\documentclass[%
 reprint,
 amsmath,amssymb,
 aps,
 pra,
]{revtex4-2}
\usepackage{graphicx}
\usepackage[]{hyperref}
\hypersetup{
    colorlinks,
    linkcolor=blue,
    citecolor=blue,
    urlcolor=blue
}
\usepackage{breakurl}
\usepackage{dcolumn}
\usepackage{bm}

\begin{document}

\preprint{APS/123-QED}

\title{Measurement of the $s$-wave scattering length between metastable helium isotopes}
\author{S. Kannan$^1$}
\author{Y. S. Athreya$^1$}
\author{A. H. Abbas$^{1}$}
\altaffiliation[Present address: ]{Artificial Intelligence and Cyber Futures Institute, Charles Sturt University, Bathurst, NSW 2795, Australia.}
\author{X. T. Yan}
\author{S. S. Hodgman$^1$}
\author{A. G. Truscott$^1$}
 \email{andrew.truscott@anu.edu.au}
\affiliation{$^1$Department of Quantum Science and Technology, Research School of Physics, The Australian National University, Canberra, ACT 2601, Australia.
}%




\date{\today}

\begin{abstract}
We report the first experimental determination of the interspecies $s$-wave scattering length\,($a_{34}$) between the $2\,^3S_1\,(F=3/2,m_F=3/2)$ state of $^3$He$^*$ and the $2\,^3S_1\,(m_J=1)$ state of $^4$He$^*$. We determine $a_{34}$ by inducing oscillations in a trapped Bose-Einstein condensate of $^4$He$^*$ and measuring the damping rate of these oscillations due to the presence of $^3$He$^*$ atoms. The deduced value of $a_{34}=29\pm3$\,nm is in good agreement with theoretical predictions. The knowledge of this scattering length is important for many fundamental experiments between these helium isotopes.
\end{abstract}
\maketitle
\emph{Introduction.} Scattering between two atoms can take many forms ($s$-wave, $p$-wave, $d$-wave etc), each with a specific scattering symmetry and a collision cross section commonly defined in terms of the scattering length\,\cite{sakurai2020modern}. At ultracold temperatures, $s$-wave scattering dominates. This means that the $s$-wave scattering length $a$ determines many of the key properties of degenerate atomic gases, including the stability of the system\,\cite{ruprecht1995time,santos2000bose,modugno2002collapse,koch2008stabilization}, thermalisation rate\,\cite{PhysRevB.34.3476,PhysRevLett.74.5202,Anderson,mosk2001mixture}, healing length of a condensate\,\cite{PhysRevA.55.2126,pethick2008bose} and phase transitions between different states such as the BEC-BCS crossover\,\cite{zwierlein2005vortices,hoinka2017goldstone,PhysRevX.7.041004,koch2023quantum} and the superfluid to Mott insulator\,\cite{greiner2002quantum,PhysRevA.100.013611}. This is especially important for mixtures of different atoms, where the multiple species present lead to a rich array of behaviors, including phase separated regimes\,\cite{molmer1998bose,PhysRevA.59.2974,marchetti2008phase,linder2010probing}, collective excitations\,\cite{graham1998collective,santiago2023collective}, polaron physics\,\cite{kohstall2012metastability,cetina}, ferromagnetism\,\cite{ferro2009,bresolin2023oscillating}, and a supersolid phase\,\cite{PhysRevLett.122.130405,PhysRevX.9.011051,PhysRevX.9.021012}.
\par
A mixture of particular interest comprises the two long-lived metastable isotopes of helium (He$^*$) in the first excited state, namely, fermionic $^3$He$^*$ and bosonic $^4$He$^*$. The single atom detection enabled by the high internal energy of He$^*$ uniquely positions it to perform a range of fundamental quantum atom optics experiments\,\cite{schellekens2005hanbury,jeltes2007comparison,hodgman2011direct,manning2015wheeler,lopes2015atomic}, and a mixture between the two isotopes extends such experiments further.
\par
Many such experiments, particularly those utilising $s$-wave collisional halos\,\cite{PhysRevLett.99.150405,khakimov2016ghost,hodgman2017solving,thomas2022matter} and interacting lattice physics\,\cite{greiner2002quantum,fabbri2015dynamical,tenart2020two,butera2021position,herce2021studying}, rely on knowledge of the $s$-wave scattering length. To realize the full potential that would come from extending these types of experiments to a mixture of $^3$He$^*$ and $^4$He$^*$ requires a determination of the interspecies $s$-wave scattering length $a_{34}$ ($2\,^3S_1\,(F=3/2,m_F=3/2)$ state of $^3$He$^*$ and $2\,^3S_1\,(m_J=1)$ state of $^4$He$^*$). There have been two previous theoretical studies in which $a_{34}$ was calculated: the first using an electronic structure calculation found $a_{34}$= $28.8^{+3.9}_{-3.3}$\,nm\,\cite{mcnamara2006degenerate}, while the second used a close coupled model to give a result of $a_{34}$=$29\pm4$\,nm\,\cite{hirsch2021close}. Goosen \emph{et al.}\,\cite{PhysRevA.82.042713} have also studied the Feshbach resonances for metastable helium atoms by using the asymptotic-bound-state model.
Neither \cite{hirsch2021close} or \cite{PhysRevA.82.042713} predicted an experimentally accessible Feshbach resonance between the $2\,^3S_1\,(F=3/2,m_F=3/2)$ state of $^3$He$^*$ and the $2\,^3S_1\,(m_J=1)$ state of $^4$He$^*$ (the most relevant experimental states, as they suppress losses due to Penning ionization \,\cite{mcnamara2006degenerate} and can be magnetically trapped), meaning that $a_{34}$ is expected to be constant at all magnetic fields. To date, there has been no corresponding experimental measurement of $a_{34}$ to test these predictions, although the observed efficiency of evaporative cooling between $^3$He$^*$ and $^4$He$^*$\,\cite{thomas2023production} plus the stability of the mixture is consistent with a relatively large and positive value.
\par
In this work, we present an innovative method for measuring the scattering length between two species in a harmonic trap via trap oscillations and out-coupling (pulsed atom laser)\,\cite{manning2010hanbury,henson2022measurement,henson2022trap}. The rest of the paper is organized as follows. First, we discuss the model and method for calculating $a_{34}$ then we present our result and address the possible source of errors. Finally, we present our conclusions and future experiments possible due to this large positive scattering length.

\begin{figure}
\centering
\includegraphics[width=5.30cm, height=6.5cm,trim=0cm 0.0cm 0.0cm 0cm, clip=true]{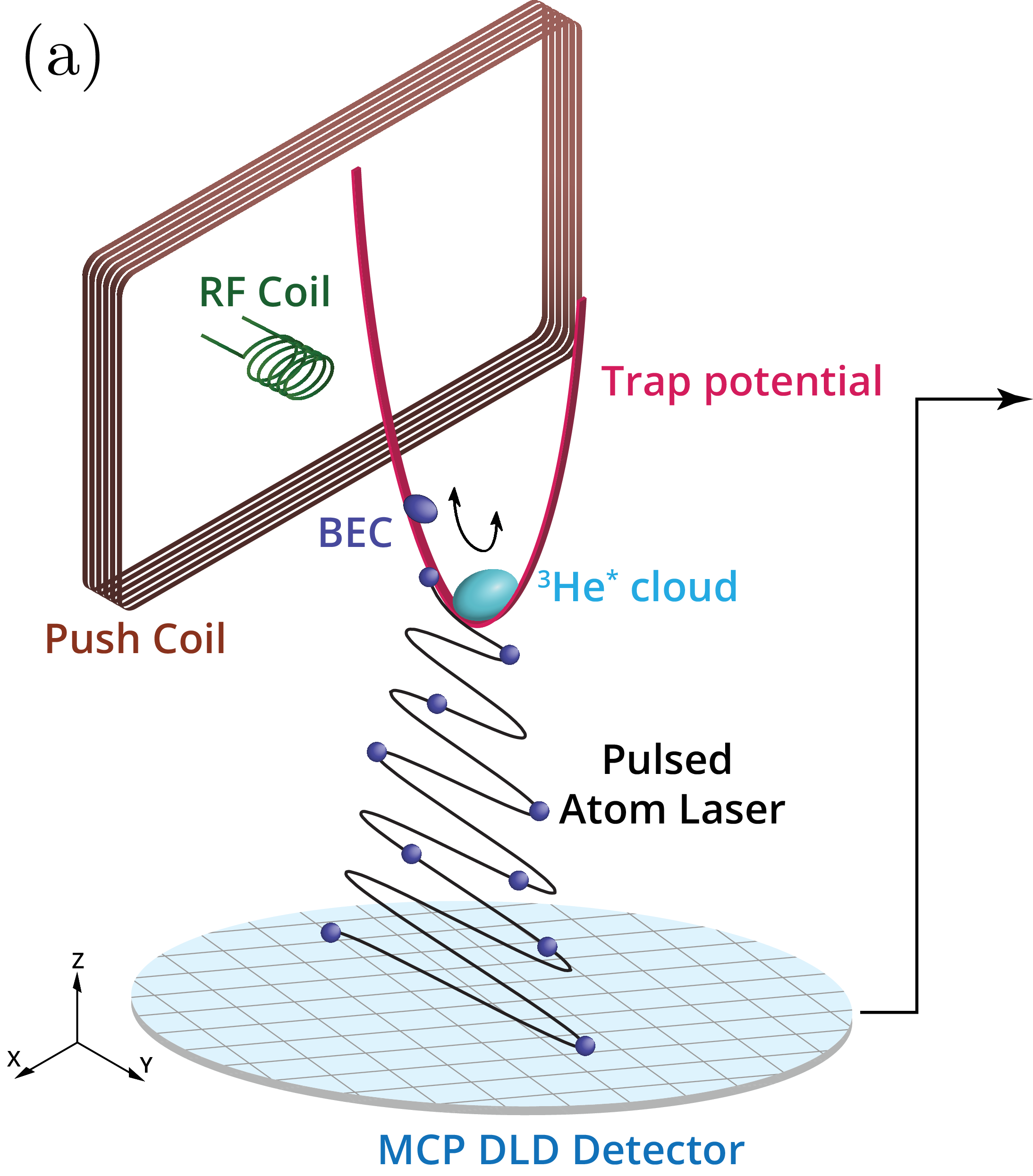}
\includegraphics[width=3.25cm, height=6.35cm]{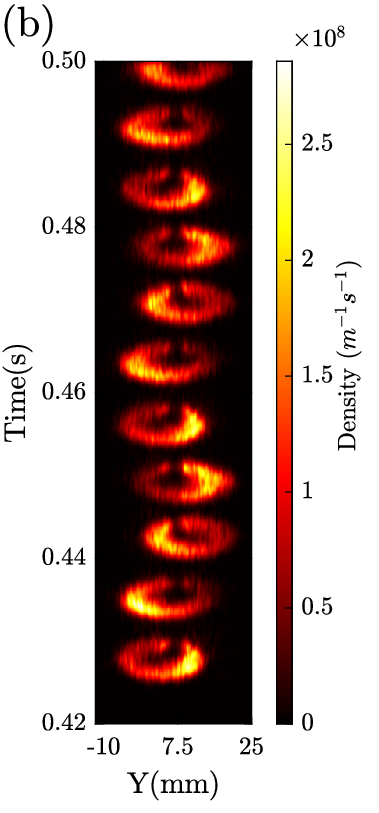}
\caption{\label{fig1}(a) Experimental schematic. A BEC of $^4$He$^*$ atoms (indigo sphere) in a harmonic magnetic trap (red parabola) is made to oscillate along the $y$-axis via a short current pulse through a `push' coil (brown coils). The velocity of the oscillating BEC is periodically sampled by radio frequency pulses produced using an RF coil to transfer a small fraction of atoms into the untrapped state. The untrapped atoms form a train of pulses that fall onto the MCP-DLD detector placed below the trap. As shown in (b) for a sample pulse train, the measured mean position of each pulse at the detector gives their in-trap velocity over time.  In the presence of a Fermi gas of $^3$He$^*$ atoms (blue sphere), collisions between the two species will lead to a damping of the oscillations.  This damping rate is dependent on the inter-species scattering length $a_{34}$.}
\end{figure}

\emph{Model and Methodology.} The starting point for our experiments is a degenerate Bose-Fermi mixture of metastable helium atoms, magnetically trapped in the bi-planar quadrupole Ioffe configuration (BiQUIC) trap\,\cite{dall2007bose}\,(Fig.\,\ref{fig1}(a)). A Bose-Einstein condensate\,(BEC) of $^4$He$^*$ is achieved via Doppler cooling and forced RF evaporation in the $m_J=+1$ sublevel of the $2\,^3S_1$ state. An ultracold cloud of $^3$He$^*$ is realised by sympathetic cooling with $^4$He$^*$\,\cite{mcnamara2006degenerate,thomas2023production} in the $m_F=+3/2$ sublevel of the $2\,^3S_1,\,F=3/2$ hyperfine state. By changing the detuning of the $^3$He$^*$ cooling laser, we are able to vary the ratio of the number of atoms in each of the two species $(N_{3}$/$N_{4})$ from 0 to 0.2. 
Once the evaporative cooling and rethermalisation have finished, we apply a short ($\sim50\,\mu$s) pulse of current to a coil oriented so as to create a momentary magnetic field gradient along the $y$-axis. This pulse sets the BEC into oscillation in the $y$-axis at the frequency of the magnetic trap $\omega_y$\,(see Fig.\,\ref{fig1}(a)).
\par
\begin{figure}
\centering
\includegraphics[width=8.5cm, height=5.8cm,trim=0.0cm 0.0cm 0.0cm 0.0cm, clip=true]{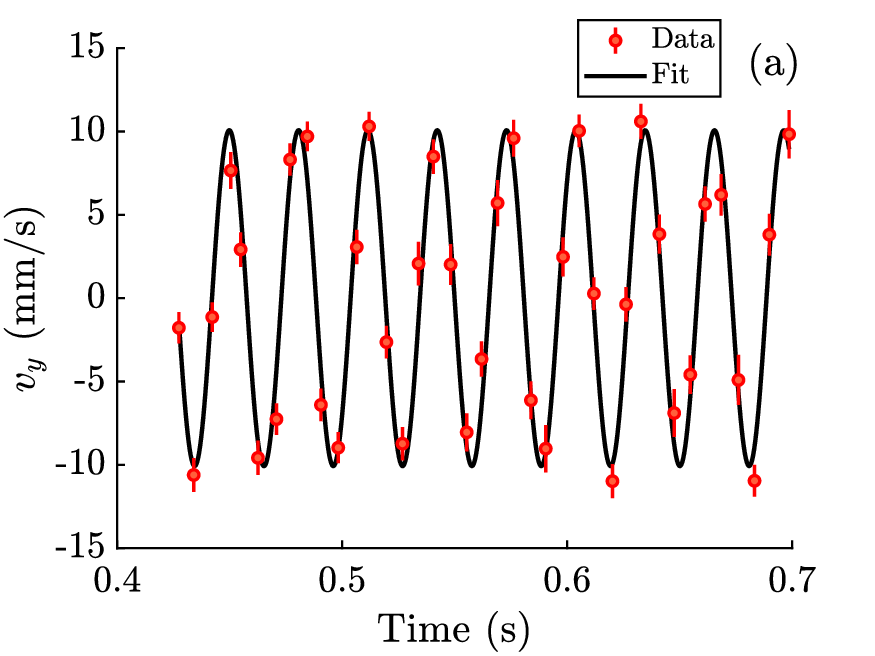}
\includegraphics[width=8.5cm, height=5.8cm,trim=0.0cm 0.0cm 0.0cm 0.0cm, clip=true]{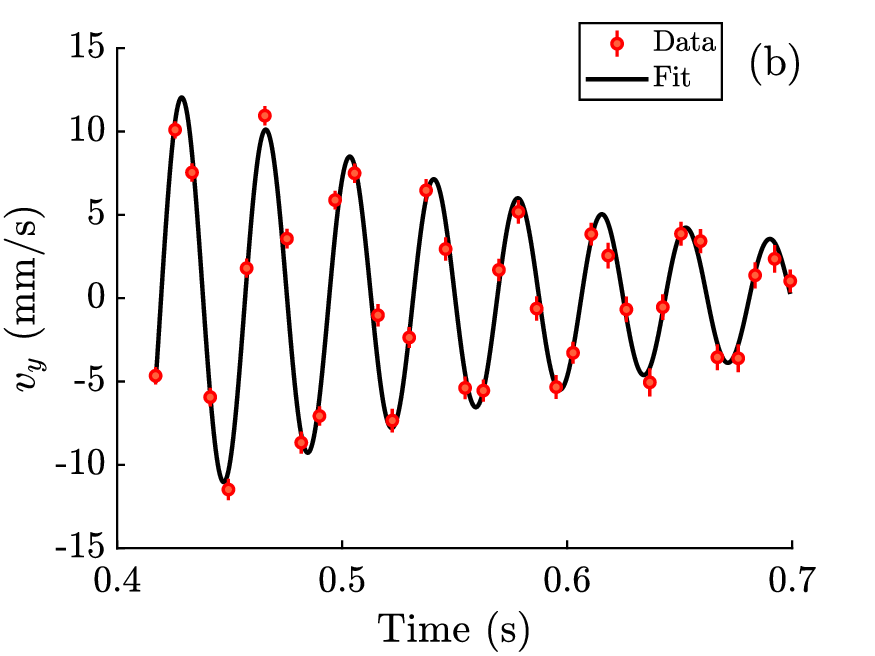}
\caption{\label{fig2}The reconstructed mean intrap velocity\,$(v_y)$ of the BEC in the magnetic trap. Red circles show experimentally determined $v_y$ values and the error bar is the standard deviation from the Gaussian fit of the pulse, while the solid lines show a fit to Eq.\,\eqref{eq1}. (a) When only a BEC is present in the trap $(2.0(5)\times10^4$\,atoms$)$, there is minimal damping and the damping rate obtained from the fit is $0.002(1)$\,s$^{-1}$. (b) In contrast, when both a pure BEC\,$(3.0(3)\times10^4$\,atoms$)$, with no detectable thermal fraction and $^3$He$^*$\,$(1.0(2)\times10^3$\,atoms$)$ are present the oscillations damp much more quickly. The corresponding damping rate is $4.6(5)$\,s$^{-1}$.}
\end{figure}

To measure the BEC oscillations and their damping rate, we out-couple a series of pulses, each containing a small fraction of $^4$He$^*$ atoms ($\sim 1\%$ of total BEC atoms per pulse). This is achieved by transferring atoms from the trapped state ($m_J=+1$) to the untrapped state ($m_J=0$) using RF pulses, forming a pulsed atom laser\,\cite{manning2010hanbury}\,(Fig.\,\ref{fig1}(a)). $^3$He$^*$ atoms are unaffected by the RF, as they are not resonant with the applied radiation. The out-coupled atoms fall 850\,mm and are detected using a micro-channel plate (MCP) and delay line detector (DLD) system\,\cite{manning2010hanbury}. As this detector is essentially in the far field, the in-trap dynamics are reconstructed
from the mean velocity of the condensate
over time, which is inferred from the out-coupled pulses' positional and temporal information on the detector. A train of these pulses for an experimental run is shown in Fig.\,\ref{fig1}(b). Note that the shape of these atom laser pulses is due to their interaction with the condensate and is well understood\,\cite{PhysRevA.97.063601}. The mean velocity for each of these pulses is extracted by fitting a Gaussian to the position of the cloud at the detector. A damped sine wave is then used to fit the oscillation of these mean velocities\,\cite{henson2022trap} 
\begin{equation}\label{eq1}
    v_y(t)=Ae^{-\gamma_4 t}\omega_a\sin(\omega_at+\phi),
\end{equation} 
with the amplitude ($A$), damping rate ($\gamma_4$), apparent frequency ($\omega_a$), and the phase ($\phi$) as free fit parameters.
In our experiment, the trap frequency in the oscillation axis is measured to be $\omega_y=2\pi\times\,1060.8(1)$\,Hz. But due to aliasing (arising because the time between the RF pulses (8.3\,ms) used for outcoupling is greater than $1/\omega_y$), we measure an apparent frequency smaller than the trap frequency (Fig.\,\ref{fig2}). We determine the true frequency of the harmonic trap by Nyquist zone fitting (for more details see Section\,3 in Ref.\,\cite{henson2022trap}).
\par
When there are no $^3$He$^*$ atoms present, the superfluid BEC oscillates without friction and there is very little damping to the BEC oscillation, as shown in Fig.\,\ref{fig2}(a).  
This non-zero damping rate can be attributed to the presence of a small fraction of thermal atoms along with the condensate which results in Landau damping\,\cite{fedichev1998damping}.  
The damping rate is increased in the presence of a $^3$He$^*$ cloud, due to $s$-wave collisions between the two species. These collisions and subsequent thermalization reduce the oscillations in a manner that depends on the $s$-wave scattering length.  An example of this damping is shown in Fig.\,\ref{fig2}(b).  
To understand the damping and thus use it to compute the scattering length $a_{34}$, we model the damped oscillation using a similar approach as in\,\cite{ferrari2002collisional}.
The dynamics of both species in a harmonic trap are analyzed by looking at both species' center of mass motion. The corresponding equation of motion along the $y$ direction for the center of mass ($y$) of a BEC in a 1D-harmonic trap containing a $^3$He$^*$ atoms is described by\,\cite{ferrari2002collisional}
\begin{equation}\label{eq2}
    \Ddot{y}=-\omega_y^2 y-\frac{4}{3}\frac{m_3N_3}{(m_3+m_4)(N_3+N_4)}\Gamma \Dot{y}.
\end{equation}
The first term is the acceleration term in the harmonic trap with $\omega_y$ as the trapping frequency along the $y$-axis, and the second term is the acceleration from the momentum kicks imparted due to instantaneous $s$-wave collisions. $m_3\,(m_4)$ is the mass of $^3$He$^*$\,($^4$He$^*$) respectively, $N_3\,(N_4)$ is the number of $^3$He$^*$\,($^4$He$^*$) in the trap, and $\Gamma$ is the instantaneous $s$-wave collision rate given by $\Gamma=\Bar{n}\sigma v_{rel}$, where $\Bar{n}$ is the effective density of $^3$He$^*$ and $^4$He$^*$\,\cite{ferrari2002collisional} \begin{equation}
    \Bar{n}=\Big(\frac{1}{N_3}+\frac{1}{N_4}\Big)\int n_3(\Vec{x})n_4(\Vec{x})d^3\Vec{x},
\end{equation} and $n_3\,(n_4)$ is the number density of $^3$He$^*$\,($^4$He$^*$). The volume of the BEC is calculated from the Thomas-Fermi radius and that of the $^3$He$^*$ cloud from the spatial density distribution of fermionic cloud\,\cite{molmer1998bose}. $\sigma=4\pi a_{34}^2$ is the scattering cross section and $v_{rel}$ is the rms relative velocity.

\emph{Results.} In this section, we discuss the calculation of $s$-wave scattering length from the damping rate $\gamma_4$ of the BEC oscillation. 
For our system, we assume that the center of mass velocity of the $^3$He$^*$ cloud is zero. We can see why this is true by looking at the damping rate $\gamma_4$  of the BEC in a $^3$He$^*$ reservoir. As the solution of the equation of motion given in Eq.\,\eqref{eq2} is the damped sine wave in Eq.\,\eqref{eq1}, the $\gamma_4$ obtained from the BEC oscillation can be expressed as
\begin{equation}\label{eq4}
  \gamma_4=\frac{2m_3N_3}{3(m_3+m_4)(N_3+N_4)}\Bar{n}\sigma v_{rel}.
\end{equation}
Now, similarly, if we consider the motion of the $^3$He$^*$ cloud, the corresponding damping rate $\gamma_3$ of its oscillation gives
\begin{equation}\label{eq5}
    \gamma_3=\frac{4}{3}\times\frac{N_4}{N_3}\times\gamma_4.
\end{equation}
In our case, $N_4 \gg N_3$, which makes the damping rate $\gamma_3$ very high relative to $\gamma_4$, and thus the approximation that the $^3$He$^*$ cloud is at rest will be valid (except at very short time scales). 
Also, note that the BEC is always immersed in the Fermi gas which has a larger radius in our experiment, and the BEC oscillation amplitude is less than this radius making the overlap between the two species very large throughout the oscillations.
\begin{figure}[ht]
\centering
\includegraphics[width=8.8cm, height=6.5cm]{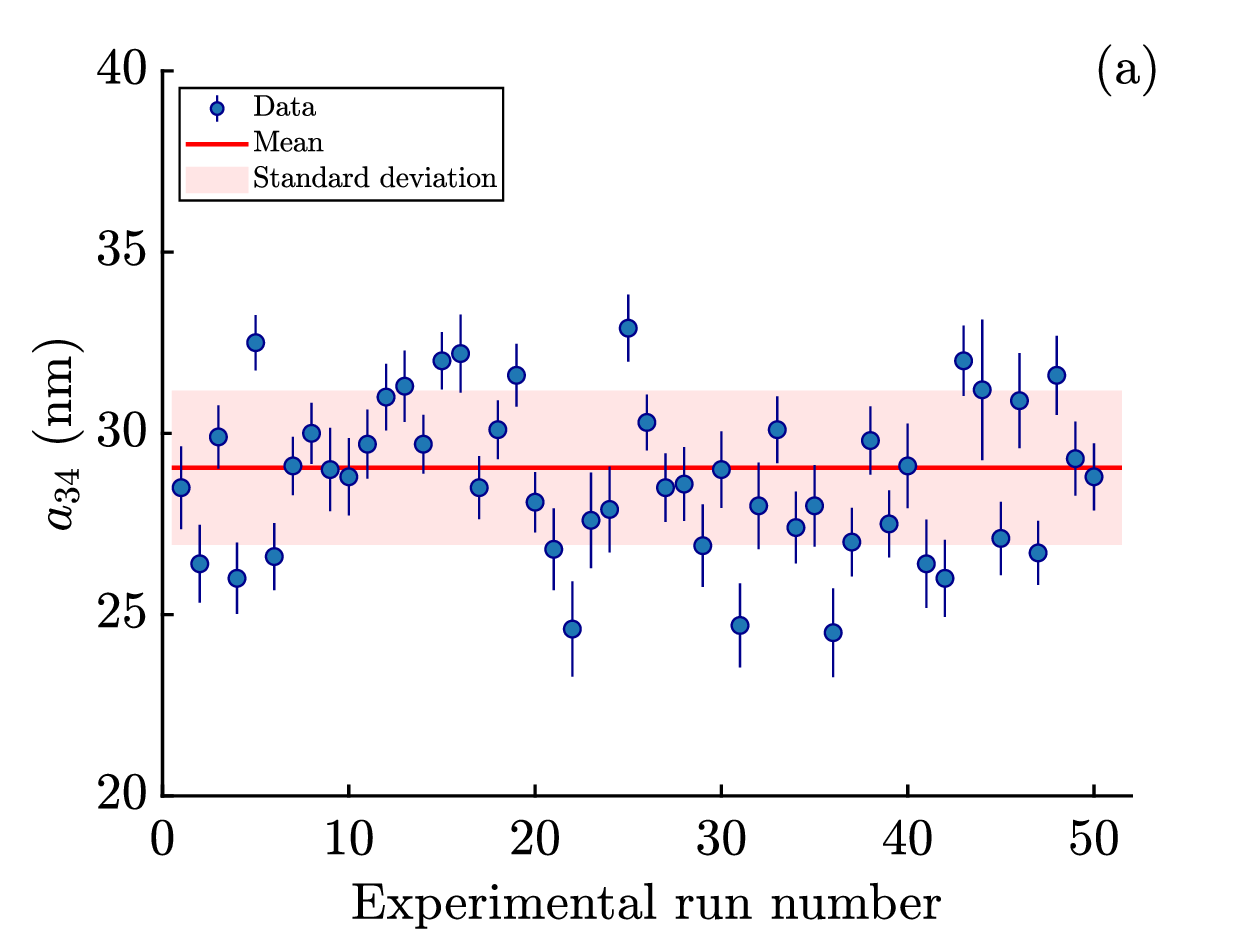}
\includegraphics[width=8.8cm, height=6.5cm]{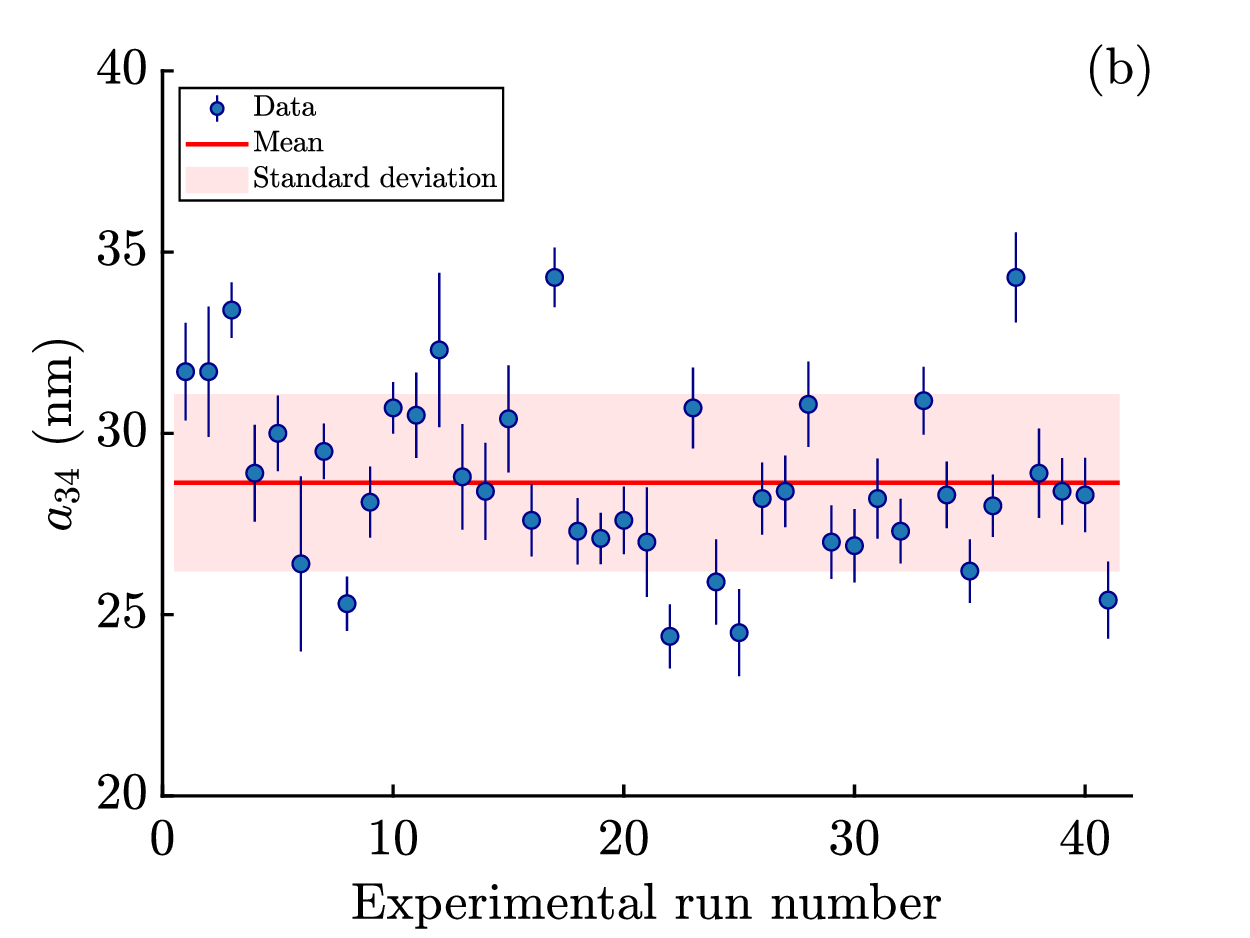}
sea\caption{\label{fig3} Scattering length obtained for different experimental runs in two different traps. The traps are (a) $\omega_{x,y,z}=2\pi\times(62.70(5),658.5(1),660.0(1))$ Hz and (b) $\omega_{x,y,z}=2\pi\times(62.77(5),1060.8(1),1062.5(1))$ Hz respectively. The horizontal red line indicates the mean value of the measurement, and the red shaded region is 1 standard deviation based on the average of all the measured $a_{34}$. The extracted scattering length from the two traps are (a) $a_{34}=29\pm3$\,nm and (b) $a_{34}=29\pm2$\,nm respectively.}
\end{figure}
\par
We extract the $s$-wave scattering length from the damping rate $\gamma_4$ in Eq.\,\eqref{eq4}, using the experimental parameters such as the number of $^4$He$^*$ atoms in the BEC $N_4$ and the number of $^3$He$^*$ atoms $N_3$. We measure $N_4$ by out-coupling all the $^4$He$^*$ atoms in the BEC as a pulsed atom laser. Each pulse (see e.g.\,Fig.\,\ref{fig1}(b)) contains a small percentage of the total BEC atoms. Adding up all these pulses and accounting for the quantum efficiency of the MCP ($0.20(2)$), measured using the procedure in \cite{PhysRevLett.105.190402}) allows us to determine the total number of $^4$He$^*$ atoms in every experimental run. 
Out-coupling a small fraction in multiple pulses, as well as allowing the direct measurement of the trap oscillations in a single shot, avoids any problems due to detector saturation that would occur at high count rates for MCP. The detuning of the $^3$He$^*$ cooling laser is set to ensure that both $N_3$ and the ratio $N_3/N_4$ remain small in all our experimental runs. This is to keep $\gamma_3$ large (see Eq.\,\eqref{eq5}) while the overall $\gamma_4$ is relatively small, as well as to cause any oscillation of the $^3$He$^*$ cloud to damp out quickly. This also allows us to directly determine $N_3$ by letting the $^3$He$^*$ cloud fall onto the detector at the end of the experimental sequence and counting the atoms measured (again, correcting for MCP quantum efficiency). The small number of atoms in the $^3$He$^*$ cloud ensures that there is negligible saturation of the detector. 
\begin{figure}[t]
\includegraphics[width=8.8cm, height=6.5cm]{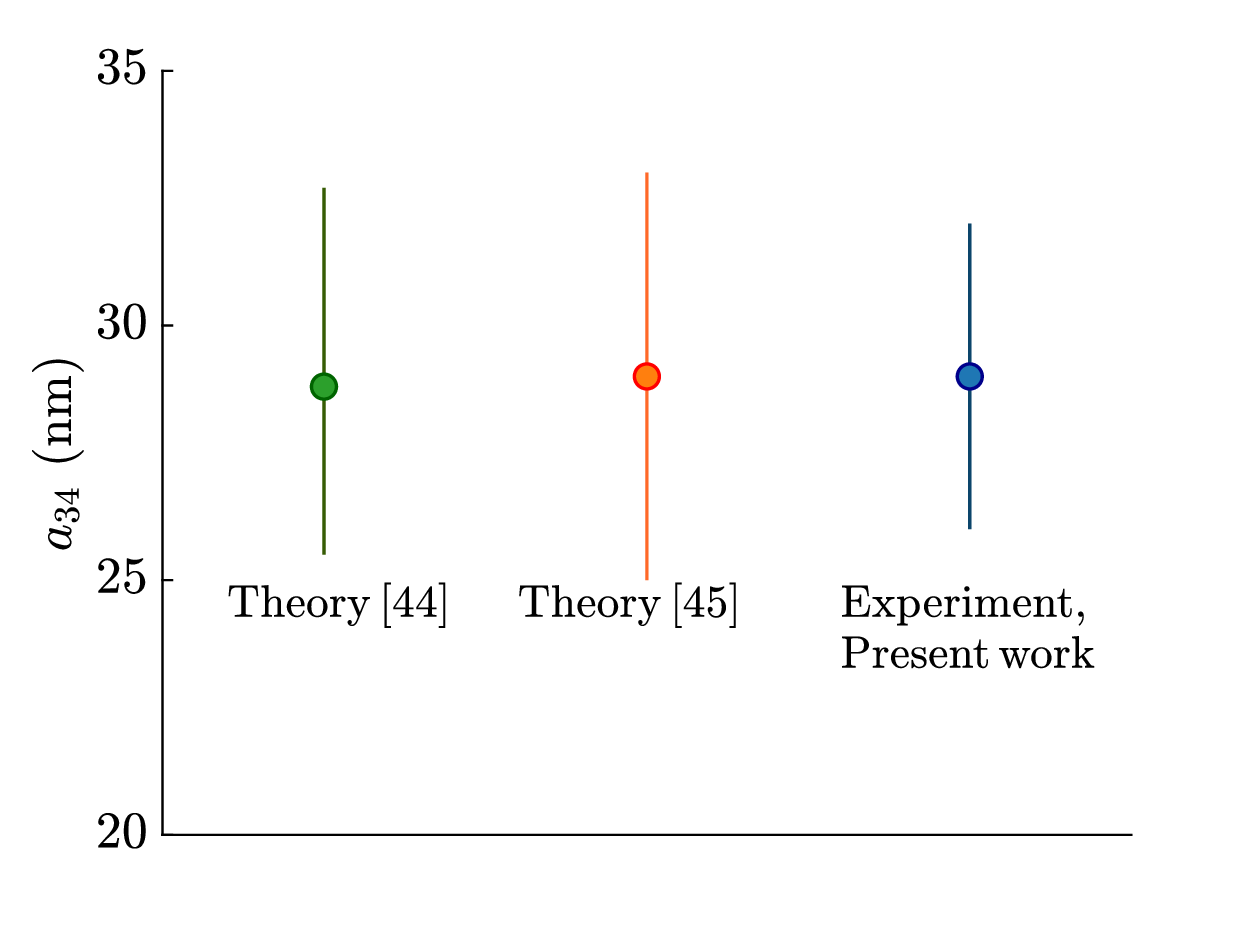}
\caption{\label{fig4} Comparison of the present result with previous determinations (all theoretical) of the $s$-wave scattering length between $^3$He$^*$ and $^4$He$^*$: from the electronic structure calculation\,\cite{mcnamara2006degenerate} ($a_{34}=28.8^{+3.9}_{-3.3}$\,nm), and from the close-coupled model of a Feshbach resonance\,\cite{hirsch2021close} ($a_{34}=29\pm4$\,nm). Our result shows $a_{34}=29\pm3$\,nm, which compares well within uncertainties with the theoretical results.}
\end{figure}
\par
Throughout the experiment, we vary the $^3$He$^*$ number from 250(30) to 1000(100) and the BEC number from 2000(200) to 20,000(2000) in different experimental runs to extract the scattering length. We note here that no correlations were observed in the measured scattering length with the number of BEC and $^3$He$^*$ atoms. 
In each experimental run, we measure all these quantities and obtain $a_{34}$ accounting for shot noise errors due to BEC and $^3$He$^*$ atom numbers and also the error in the fitting of the damping rate $\gamma_4$. The plots of extracted scattering length for two different magnetic traps $\omega_{x,y,z}=2\pi\times(62.70(5),658.5(1),660.0(1))$ Hz and $\omega_{x,y,z}=2\pi\times(62.77(5),1060.8(1),1062.5(1))$ Hz are given in Fig.\,\ref{fig3}. Two different traps were used to check if there was any dependence between the extracted scattering length and the trapping frequencies, as $\omega$ affects the atomic density. The result (Fig.\,\ref{fig3}) shows that these quantities are uncorrelated. We report the final measure of $a_{34}$ by combining all the data in the two traps and calculating the statistical and systematic errors in the experiment gives
\begin{equation}\label{eq6}
a_{34}=29\pm2\,\text{(stat)}\pm1\,\text{(syst)}\,\text{nm}.
\end{equation}
The uncertainty in the quantum efficiency of the detector is the largest source of systematic error in our measurement. We also note that though there are 2-body losses, and a boson-boson-fermion (BBF) 3-body loss rate in the mixture\,\cite{mcnamara2006degenerate}, the small density of the two species for the experiment results in a low number loss in our time frame and hence this would only be a minor correction to our model.
\par
Our result agrees well with the theoretical predictions of $a_{34}$ (see Fig.\,\ref{fig4})
and has comparable uncertainties. The uncertainty in the determination of $N_3$ and $N_4$ gives rise to systematic errors and this can be reduced by using a detector with a smaller uncertainty in the quantum efficiency.

\emph{Conclusion.} To summarise, we have used a method similar to that in \cite{ferrari2002collisional}, consisting of trap oscillations and out-coupling to measure the interspecies $s$-wave scattering length between the $2\,^3S_1\,(F=3/2,m_F=3/2)$ state of $^3$He$^*$ and the $2\,^3S_1\,(m_J=1)$ state of $^4$He$^*$. We obtain a value of the $s$-wave scattering length to be $29\pm3$\,nm, which is the first experimental determination of this parameter. Our measured value is also in good agreement with existing theoretical predictions (Fig.\,\ref{fig4}). Furthermore, this large positive value paves the way for collision experiments between the species, such as a demonstration of non-locality using a Bell inequality with a momentum entangled state\,\cite{thomas2022matter,PhysRevA.91.052114} of $^3$He$^*$ and $^4$He$^*$ atoms or testing the validity of the weak equivalence principle for these nonclassical states\,\cite{PhysRevLett.120.043602}.

\emph{Acknowledgments.} This work was supported through Australian Research
Council Discovery Projects Grant No. DP190103021, DP240101346 and DP240101441. S.S.H.
was supported by Australian Research Council Future Fellowship
Grant No. FT220100670. S. K. was supported
by an Australian Government Research Training Program
scholarship.


\bibliography{apssamp}

\providecommand{\noopsort}[1]{}\providecommand{\singleletter}[1]{#1}%
\begin{thebibliography}{57}%
\makeatletter
\providecommand \@ifxundefined [1]{%
 \@ifx{#1\undefined}
}%
\providecommand \@ifnum [1]{%
 \ifnum #1\expandafter \@firstoftwo
 \else \expandafter \@secondoftwo
 \fi
}%
\providecommand \@ifx [1]{%
 \ifx #1\expandafter \@firstoftwo
 \else \expandafter \@secondoftwo
 \fi
}%
\providecommand \natexlab [1]{#1}%
\providecommand \enquote  [1]{``#1''}%
\providecommand \bibnamefont  [1]{#1}%
\providecommand \bibfnamefont [1]{#1}%
\providecommand \citenamefont [1]{#1}%
\providecommand \href@noop [0]{\@secondoftwo}%
\providecommand \href [0]{\begingroup \@sanitize@url \@href}%
\providecommand \@href[1]{\@@startlink{#1}\@@href}%
\providecommand \@@href[1]{\endgroup#1\@@endlink}%
\providecommand \@sanitize@url [0]{\catcode `\\12\catcode `\$12\catcode `\&12\catcode `\#12\catcode `\^12\catcode `\_12\catcode `\%12\relax}%
\providecommand \@@startlink[1]{}%
\providecommand \@@endlink[0]{}%
\providecommand \url  [0]{\begingroup\@sanitize@url \@url }%
\providecommand \@url [1]{\endgroup\@href {#1}{\urlprefix }}%
\providecommand \urlprefix  [0]{URL }%
\providecommand \Eprint [0]{\href }%
\providecommand \doibase [0]{https://doi.org/}%
\providecommand \selectlanguage [0]{\@gobble}%
\providecommand \bibinfo  [0]{\@secondoftwo}%
\providecommand \bibfield  [0]{\@secondoftwo}%
\providecommand \translation [1]{[#1]}%
\providecommand \BibitemOpen [0]{}%
\providecommand \bibitemStop [0]{}%
\providecommand \bibitemNoStop [0]{.\EOS\space}%
\providecommand \EOS [0]{\spacefactor3000\relax}%
\providecommand \BibitemShut  [1]{\csname bibitem#1\endcsname}%
\let\auto@bib@innerbib\@empty
\bibitem [{\citenamefont {Sakurai}\ and\ \citenamefont {Napolitano}(2020)}]{sakurai2020modern}%
  \BibitemOpen
  \bibfield  {author} {\bibinfo {author} {\bibfnamefont {J.~J.}\ \bibnamefont {Sakurai}}\ and\ \bibinfo {author} {\bibfnamefont {J.}~\bibnamefont {Napolitano}},\ }\href {https://doi.org/https://doi.org/10.1017/9781108587280} {\emph {\bibinfo {title} {Modern quantum mechanics}}}\ (\bibinfo  {publisher} {Cambridge University Press},\ \bibinfo {year} {2020})\BibitemShut {NoStop}%
\bibitem [{\citenamefont {Ruprecht}\ \emph {et~al.}(1995)\citenamefont {Ruprecht}, \citenamefont {Holland}, \citenamefont {Burnett},\ and\ \citenamefont {Edwards}}]{ruprecht1995time}%
  \BibitemOpen
  \bibfield  {author} {\bibinfo {author} {\bibfnamefont {P.~A.}\ \bibnamefont {Ruprecht}}, \bibinfo {author} {\bibfnamefont {M.~J.}\ \bibnamefont {Holland}}, \bibinfo {author} {\bibfnamefont {K.}~\bibnamefont {Burnett}},\ and\ \bibinfo {author} {\bibfnamefont {M.}~\bibnamefont {Edwards}},\ }\href {https://doi.org/10.1103/PhysRevA.51.4704} {\bibfield  {journal} {\bibinfo  {journal} {Physical Review A}\ }\textbf {\bibinfo {volume} {51}},\ \bibinfo {pages} {4704} (\bibinfo {year} {1995})}\BibitemShut {NoStop}%
\bibitem [{\citenamefont {Santos}\ \emph {et~al.}(2000)\citenamefont {Santos}, \citenamefont {Shlyapnikov}, \citenamefont {Zoller},\ and\ \citenamefont {Lewenstein}}]{santos2000bose}%
  \BibitemOpen
  \bibfield  {author} {\bibinfo {author} {\bibfnamefont {L.}~\bibnamefont {Santos}}, \bibinfo {author} {\bibfnamefont {G.~V.}\ \bibnamefont {Shlyapnikov}}, \bibinfo {author} {\bibfnamefont {P.}~\bibnamefont {Zoller}},\ and\ \bibinfo {author} {\bibfnamefont {M.}~\bibnamefont {Lewenstein}},\ }\href {https://doi.org/10.1103/PhysRevLett.85.1791} {\bibfield  {journal} {\bibinfo  {journal} {Physical Review Letters}\ }\textbf {\bibinfo {volume} {85}},\ \bibinfo {pages} {1791} (\bibinfo {year} {2000})}\BibitemShut {NoStop}%
\bibitem [{\citenamefont {Modugno}\ \emph {et~al.}(2002)\citenamefont {Modugno}, \citenamefont {Roati}, \citenamefont {Riboli}, \citenamefont {Ferlaino}, \citenamefont {Brecha},\ and\ \citenamefont {Inguscio}}]{modugno2002collapse}%
  \BibitemOpen
  \bibfield  {author} {\bibinfo {author} {\bibfnamefont {G.}~\bibnamefont {Modugno}}, \bibinfo {author} {\bibfnamefont {G.}~\bibnamefont {Roati}}, \bibinfo {author} {\bibfnamefont {F.}~\bibnamefont {Riboli}}, \bibinfo {author} {\bibfnamefont {F.}~\bibnamefont {Ferlaino}}, \bibinfo {author} {\bibfnamefont {R.~J.}\ \bibnamefont {Brecha}},\ and\ \bibinfo {author} {\bibfnamefont {M.}~\bibnamefont {Inguscio}},\ }\href {https://doi.org/10.1126/science.1077386} {\bibfield  {journal} {\bibinfo  {journal} {Science}\ }\textbf {\bibinfo {volume} {297}},\ \bibinfo {pages} {2240} (\bibinfo {year} {2002})}\BibitemShut {NoStop}%
\bibitem [{\citenamefont {Koch}\ \emph {et~al.}(2008)\citenamefont {Koch}, \citenamefont {Lahaye}, \citenamefont {Metz}, \citenamefont {Fr{\"o}hlich}, \citenamefont {Griesmaier},\ and\ \citenamefont {Pfau}}]{koch2008stabilization}%
  \BibitemOpen
  \bibfield  {author} {\bibinfo {author} {\bibfnamefont {T.}~\bibnamefont {Koch}}, \bibinfo {author} {\bibfnamefont {T.}~\bibnamefont {Lahaye}}, \bibinfo {author} {\bibfnamefont {J.}~\bibnamefont {Metz}}, \bibinfo {author} {\bibfnamefont {B.}~\bibnamefont {Fr{\"o}hlich}}, \bibinfo {author} {\bibfnamefont {A.}~\bibnamefont {Griesmaier}},\ and\ \bibinfo {author} {\bibfnamefont {T.}~\bibnamefont {Pfau}},\ }\href {https://doi.org/https://doi.org/10.1038/nphys887} {\bibfield  {journal} {\bibinfo  {journal} {Nature physics}\ }\textbf {\bibinfo {volume} {4}},\ \bibinfo {pages} {218} (\bibinfo {year} {2008})}\BibitemShut {NoStop}%
\bibitem [{\citenamefont {Hess}(1986)}]{PhysRevB.34.3476}%
  \BibitemOpen
  \bibfield  {author} {\bibinfo {author} {\bibfnamefont {H.~F.}\ \bibnamefont {Hess}},\ }\href {https://doi.org/10.1103/PhysRevB.34.3476} {\bibfield  {journal} {\bibinfo  {journal} {Physical Review B}\ }\textbf {\bibinfo {volume} {34}},\ \bibinfo {pages} {3476} (\bibinfo {year} {1986})}\BibitemShut {NoStop}%
\bibitem [{\citenamefont {Davis}\ \emph {et~al.}(1995)\citenamefont {Davis}, \citenamefont {Mewes}, \citenamefont {Joffe}, \citenamefont {Andrews},\ and\ \citenamefont {Ketterle}}]{PhysRevLett.74.5202}%
  \BibitemOpen
  \bibfield  {author} {\bibinfo {author} {\bibfnamefont {K.~B.}\ \bibnamefont {Davis}}, \bibinfo {author} {\bibfnamefont {M.-O.}\ \bibnamefont {Mewes}}, \bibinfo {author} {\bibfnamefont {M.~A.}\ \bibnamefont {Joffe}}, \bibinfo {author} {\bibfnamefont {M.~R.}\ \bibnamefont {Andrews}},\ and\ \bibinfo {author} {\bibfnamefont {W.}~\bibnamefont {Ketterle}},\ }\href {https://doi.org/10.1103/PhysRevLett.74.5202} {\bibfield  {journal} {\bibinfo  {journal} {Physical Review Letters}\ }\textbf {\bibinfo {volume} {74}},\ \bibinfo {pages} {5202} (\bibinfo {year} {1995})}\BibitemShut {NoStop}%
\bibitem [{\citenamefont {Anderson}\ \emph {et~al.}(1995)\citenamefont {Anderson}, \citenamefont {Ensher}, \citenamefont {Matthews}, \citenamefont {Wieman},\ and\ \citenamefont {Cornell}}]{Anderson}%
  \BibitemOpen
  \bibfield  {author} {\bibinfo {author} {\bibfnamefont {M.~H.}\ \bibnamefont {Anderson}}, \bibinfo {author} {\bibfnamefont {J.~R.}\ \bibnamefont {Ensher}}, \bibinfo {author} {\bibfnamefont {M.~R.}\ \bibnamefont {Matthews}}, \bibinfo {author} {\bibfnamefont {C.~E.}\ \bibnamefont {Wieman}},\ and\ \bibinfo {author} {\bibfnamefont {E.~A.}\ \bibnamefont {Cornell}},\ }\href {https://doi.org/10.1126/science.269.5221.198} {\bibfield  {journal} {\bibinfo  {journal} {Science}\ }\textbf {\bibinfo {volume} {269}},\ \bibinfo {pages} {198} (\bibinfo {year} {1995})}\BibitemShut {NoStop}%
\bibitem [{\citenamefont {Mosk}\ \emph {et~al.}(2001)\citenamefont {Mosk}, \citenamefont {Kraft}, \citenamefont {Mudrich}, \citenamefont {Singer}, \citenamefont {Wohlleben}, \citenamefont {Grimm},\ and\ \citenamefont {Weidem{\"u}ller}}]{mosk2001mixture}%
  \BibitemOpen
  \bibfield  {author} {\bibinfo {author} {\bibfnamefont {A.}~\bibnamefont {Mosk}}, \bibinfo {author} {\bibfnamefont {S.}~\bibnamefont {Kraft}}, \bibinfo {author} {\bibfnamefont {M.}~\bibnamefont {Mudrich}}, \bibinfo {author} {\bibfnamefont {K.}~\bibnamefont {Singer}}, \bibinfo {author} {\bibfnamefont {W.}~\bibnamefont {Wohlleben}}, \bibinfo {author} {\bibfnamefont {R.}~\bibnamefont {Grimm}},\ and\ \bibinfo {author} {\bibfnamefont {M.}~\bibnamefont {Weidem{\"u}ller}},\ }\href {https://doi.org/10.1007/s003400100743} {\bibfield  {journal} {\bibinfo  {journal} {Applied Physics B}\ }\textbf {\bibinfo {volume} {73}},\ \bibinfo {pages} {791} (\bibinfo {year} {2001})}\BibitemShut {NoStop}%
\bibitem [{\citenamefont {Lundh}\ \emph {et~al.}(1997)\citenamefont {Lundh}, \citenamefont {Pethick},\ and\ \citenamefont {Smith}}]{PhysRevA.55.2126}%
  \BibitemOpen
  \bibfield  {author} {\bibinfo {author} {\bibfnamefont {E.}~\bibnamefont {Lundh}}, \bibinfo {author} {\bibfnamefont {C.~J.}\ \bibnamefont {Pethick}},\ and\ \bibinfo {author} {\bibfnamefont {H.}~\bibnamefont {Smith}},\ }\href {https://doi.org/10.1103/PhysRevA.55.2126} {\bibfield  {journal} {\bibinfo  {journal} {Physical Review A}\ }\textbf {\bibinfo {volume} {55}},\ \bibinfo {pages} {2126} (\bibinfo {year} {1997})}\BibitemShut {NoStop}%
\bibitem [{\citenamefont {Pethick}\ and\ \citenamefont {Smith}(2008)}]{pethick2008bose}%
  \BibitemOpen
  \bibfield  {author} {\bibinfo {author} {\bibfnamefont {C.~J.}\ \bibnamefont {Pethick}}\ and\ \bibinfo {author} {\bibfnamefont {H.}~\bibnamefont {Smith}},\ }\href {https://doi.org/https://doi.org/10.1017/CBO9780511802850} {\emph {\bibinfo {title} {Bose--Einstein condensation in dilute gases}}}\ (\bibinfo  {publisher} {Cambridge university press},\ \bibinfo {year} {2008})\BibitemShut {NoStop}%
\bibitem [{\citenamefont {Zwierlein}\ \emph {et~al.}(2005)\citenamefont {Zwierlein}, \citenamefont {Abo-Shaeer}, \citenamefont {Schirotzek}, \citenamefont {Schunck},\ and\ \citenamefont {Ketterle}}]{zwierlein2005vortices}%
  \BibitemOpen
  \bibfield  {author} {\bibinfo {author} {\bibfnamefont {M.~W.}\ \bibnamefont {Zwierlein}}, \bibinfo {author} {\bibfnamefont {J.~R.}\ \bibnamefont {Abo-Shaeer}}, \bibinfo {author} {\bibfnamefont {A.}~\bibnamefont {Schirotzek}}, \bibinfo {author} {\bibfnamefont {C.~H.}\ \bibnamefont {Schunck}},\ and\ \bibinfo {author} {\bibfnamefont {W.}~\bibnamefont {Ketterle}},\ }\href {https://doi.org/https://doi.org/10.1038/nature03858} {\bibfield  {journal} {\bibinfo  {journal} {Nature}\ }\textbf {\bibinfo {volume} {435}},\ \bibinfo {pages} {1047} (\bibinfo {year} {2005})}\BibitemShut {NoStop}%
\bibitem [{\citenamefont {Hoinka}\ \emph {et~al.}(2017)\citenamefont {Hoinka}, \citenamefont {Dyke}, \citenamefont {Lingham}, \citenamefont {Kinnunen}, \citenamefont {Bruun},\ and\ \citenamefont {Vale}}]{hoinka2017goldstone}%
  \BibitemOpen
  \bibfield  {author} {\bibinfo {author} {\bibfnamefont {S.}~\bibnamefont {Hoinka}}, \bibinfo {author} {\bibfnamefont {P.}~\bibnamefont {Dyke}}, \bibinfo {author} {\bibfnamefont {M.~G.}\ \bibnamefont {Lingham}}, \bibinfo {author} {\bibfnamefont {J.~J.}\ \bibnamefont {Kinnunen}}, \bibinfo {author} {\bibfnamefont {G.~M.}\ \bibnamefont {Bruun}},\ and\ \bibinfo {author} {\bibfnamefont {C.~J.}\ \bibnamefont {Vale}},\ }\href {https://doi.org/https://doi.org/10.1038/nphys4187} {\bibfield  {journal} {\bibinfo  {journal} {Nature Physics}\ }\textbf {\bibinfo {volume} {13}},\ \bibinfo {pages} {943} (\bibinfo {year} {2017})}\BibitemShut {NoStop}%
\bibitem [{\citenamefont {Horikoshi}\ \emph {et~al.}(2017)\citenamefont {Horikoshi}, \citenamefont {Koashi}, \citenamefont {Tajima}, \citenamefont {Ohashi},\ and\ \citenamefont {Kuwata-Gonokami}}]{PhysRevX.7.041004}%
  \BibitemOpen
  \bibfield  {author} {\bibinfo {author} {\bibfnamefont {M.}~\bibnamefont {Horikoshi}}, \bibinfo {author} {\bibfnamefont {M.}~\bibnamefont {Koashi}}, \bibinfo {author} {\bibfnamefont {H.}~\bibnamefont {Tajima}}, \bibinfo {author} {\bibfnamefont {Y.}~\bibnamefont {Ohashi}},\ and\ \bibinfo {author} {\bibfnamefont {M.}~\bibnamefont {Kuwata-Gonokami}},\ }\href {https://doi.org/10.1103/PhysRevX.7.041004} {\bibfield  {journal} {\bibinfo  {journal} {Physical Review X}\ }\textbf {\bibinfo {volume} {7}},\ \bibinfo {pages} {041004} (\bibinfo {year} {2017})}\BibitemShut {NoStop}%
\bibitem [{\citenamefont {Koch}\ \emph {et~al.}(2023)\citenamefont {Koch}, \citenamefont {Menon}, \citenamefont {Cuestas}, \citenamefont {Barbosa}, \citenamefont {Lutz}, \citenamefont {Fogarty}, \citenamefont {Busch},\ and\ \citenamefont {Widera}}]{koch2023quantum}%
  \BibitemOpen
  \bibfield  {author} {\bibinfo {author} {\bibfnamefont {J.}~\bibnamefont {Koch}}, \bibinfo {author} {\bibfnamefont {K.}~\bibnamefont {Menon}}, \bibinfo {author} {\bibfnamefont {E.}~\bibnamefont {Cuestas}}, \bibinfo {author} {\bibfnamefont {S.}~\bibnamefont {Barbosa}}, \bibinfo {author} {\bibfnamefont {E.}~\bibnamefont {Lutz}}, \bibinfo {author} {\bibfnamefont {T.}~\bibnamefont {Fogarty}}, \bibinfo {author} {\bibfnamefont {T.}~\bibnamefont {Busch}},\ and\ \bibinfo {author} {\bibfnamefont {A.}~\bibnamefont {Widera}},\ }\href {https://doi.org/https://doi.org/10.1038/s41586-023-06469-8} {\bibfield  {journal} {\bibinfo  {journal} {Nature}\ }\textbf {\bibinfo {volume} {621}},\ \bibinfo {pages} {723} (\bibinfo {year} {2023})}\BibitemShut {NoStop}%
\bibitem [{\citenamefont {Greiner}\ \emph {et~al.}(2002)\citenamefont {Greiner}, \citenamefont {Mandel}, \citenamefont {Esslinger}, \citenamefont {H{\"a}nsch},\ and\ \citenamefont {Bloch}}]{greiner2002quantum}%
  \BibitemOpen
  \bibfield  {author} {\bibinfo {author} {\bibfnamefont {M.}~\bibnamefont {Greiner}}, \bibinfo {author} {\bibfnamefont {O.}~\bibnamefont {Mandel}}, \bibinfo {author} {\bibfnamefont {T.}~\bibnamefont {Esslinger}}, \bibinfo {author} {\bibfnamefont {T.~W.}\ \bibnamefont {H{\"a}nsch}},\ and\ \bibinfo {author} {\bibfnamefont {I.}~\bibnamefont {Bloch}},\ }\href {https://doi.org/https://doi.org/10.1038/415039a} {\bibfield  {journal} {\bibinfo  {journal} {Nature}\ }\textbf {\bibinfo {volume} {415}},\ \bibinfo {pages} {39} (\bibinfo {year} {2002})}\BibitemShut {NoStop}%
\bibitem [{\citenamefont {Lin}\ \emph {et~al.}(2019)\citenamefont {Lin}, \citenamefont {Papariello}, \citenamefont {Molignini}, \citenamefont {Chitra},\ and\ \citenamefont {Lode}}]{PhysRevA.100.013611}%
  \BibitemOpen
  \bibfield  {author} {\bibinfo {author} {\bibfnamefont {R.}~\bibnamefont {Lin}}, \bibinfo {author} {\bibfnamefont {L.}~\bibnamefont {Papariello}}, \bibinfo {author} {\bibfnamefont {P.}~\bibnamefont {Molignini}}, \bibinfo {author} {\bibfnamefont {R.}~\bibnamefont {Chitra}},\ and\ \bibinfo {author} {\bibfnamefont {A.~U.~J.}\ \bibnamefont {Lode}},\ }\href {https://doi.org/10.1103/PhysRevA.100.013611} {\bibfield  {journal} {\bibinfo  {journal} {Physical Review A}\ }\textbf {\bibinfo {volume} {100}},\ \bibinfo {pages} {013611} (\bibinfo {year} {2019})}\BibitemShut {NoStop}%
\bibitem [{\citenamefont {M\o{}lmer}(1998)}]{molmer1998bose}%
  \BibitemOpen
  \bibfield  {author} {\bibinfo {author} {\bibfnamefont {K.}~\bibnamefont {M\o{}lmer}},\ }\href {https://doi.org/10.1103/PhysRevLett.80.1804} {\bibfield  {journal} {\bibinfo  {journal} {Physical Review Letters}\ }\textbf {\bibinfo {volume} {80}},\ \bibinfo {pages} {1804} (\bibinfo {year} {1998})}\BibitemShut {NoStop}%
\bibitem [{\citenamefont {Nygaard}\ and\ \citenamefont {M\o{}lmer}(1999)}]{PhysRevA.59.2974}%
  \BibitemOpen
  \bibfield  {author} {\bibinfo {author} {\bibfnamefont {N.}~\bibnamefont {Nygaard}}\ and\ \bibinfo {author} {\bibfnamefont {K.}~\bibnamefont {M\o{}lmer}},\ }\href {https://doi.org/10.1103/PhysRevA.59.2974} {\bibfield  {journal} {\bibinfo  {journal} {Physical Review A}\ }\textbf {\bibinfo {volume} {59}},\ \bibinfo {pages} {2974} (\bibinfo {year} {1999})}\BibitemShut {NoStop}%
\bibitem [{\citenamefont {Marchetti}\ \emph {et~al.}(2008)\citenamefont {Marchetti}, \citenamefont {Mathy}, \citenamefont {Huse},\ and\ \citenamefont {Parish}}]{marchetti2008phase}%
  \BibitemOpen
  \bibfield  {author} {\bibinfo {author} {\bibfnamefont {F.~M.}\ \bibnamefont {Marchetti}}, \bibinfo {author} {\bibfnamefont {C.~J.~M.}\ \bibnamefont {Mathy}}, \bibinfo {author} {\bibfnamefont {D.~A.}\ \bibnamefont {Huse}},\ and\ \bibinfo {author} {\bibfnamefont {M.~M.}\ \bibnamefont {Parish}},\ }\href {https://doi.org/10.1103/PhysRevB.78.134517} {\bibfield  {journal} {\bibinfo  {journal} {Physical Review B}\ }\textbf {\bibinfo {volume} {78}},\ \bibinfo {pages} {134517} (\bibinfo {year} {2008})}\BibitemShut {NoStop}%
\bibitem [{\citenamefont {Linder}\ and\ \citenamefont {Sudb{\o}}(2010)}]{linder2010probing}%
  \BibitemOpen
  \bibfield  {author} {\bibinfo {author} {\bibfnamefont {J.}~\bibnamefont {Linder}}\ and\ \bibinfo {author} {\bibfnamefont {A.}~\bibnamefont {Sudb{\o}}},\ }\href {https://doi.org/10.1103/PhysRevA.81.013622} {\bibfield  {journal} {\bibinfo  {journal} {Physical Review A}\ }\textbf {\bibinfo {volume} {81}},\ \bibinfo {pages} {013622} (\bibinfo {year} {2010})}\BibitemShut {NoStop}%
\bibitem [{\citenamefont {Graham}\ and\ \citenamefont {Walls}(1998)}]{graham1998collective}%
  \BibitemOpen
  \bibfield  {author} {\bibinfo {author} {\bibfnamefont {R.}~\bibnamefont {Graham}}\ and\ \bibinfo {author} {\bibfnamefont {D.}~\bibnamefont {Walls}},\ }\href {https://doi.org/10.1103/PhysRevA.57.484} {\bibfield  {journal} {\bibinfo  {journal} {Physical Review A}\ }\textbf {\bibinfo {volume} {57}},\ \bibinfo {pages} {484} (\bibinfo {year} {1998})}\BibitemShut {NoStop}%
\bibitem [{\citenamefont {Santiago-Garc{\'\i}a}\ and\ \citenamefont {Camacho-Guardian}(2023)}]{santiago2023collective}%
  \BibitemOpen
  \bibfield  {author} {\bibinfo {author} {\bibfnamefont {M.}~\bibnamefont {Santiago-Garc{\'\i}a}}\ and\ \bibinfo {author} {\bibfnamefont {A.}~\bibnamefont {Camacho-Guardian}},\ }\href {https://doi.org/10.1088/1367-2630/acf72d} {\bibfield  {journal} {\bibinfo  {journal} {New Journal of Physics}\ }\textbf {\bibinfo {volume} {25}},\ \bibinfo {pages} {093032} (\bibinfo {year} {2023})}\BibitemShut {NoStop}%
\bibitem [{\citenamefont {Kohstall}\ \emph {et~al.}(2012)\citenamefont {Kohstall}, \citenamefont {Zaccanti}, \citenamefont {Jag}, \citenamefont {Trenkwalder}, \citenamefont {Massignan}, \citenamefont {Bruun}, \citenamefont {Schreck},\ and\ \citenamefont {Grimm}}]{kohstall2012metastability}%
  \BibitemOpen
  \bibfield  {author} {\bibinfo {author} {\bibfnamefont {C.}~\bibnamefont {Kohstall}}, \bibinfo {author} {\bibfnamefont {M.}~\bibnamefont {Zaccanti}}, \bibinfo {author} {\bibfnamefont {M.}~\bibnamefont {Jag}}, \bibinfo {author} {\bibfnamefont {A.}~\bibnamefont {Trenkwalder}}, \bibinfo {author} {\bibfnamefont {P.}~\bibnamefont {Massignan}}, \bibinfo {author} {\bibfnamefont {G.~M.}\ \bibnamefont {Bruun}}, \bibinfo {author} {\bibfnamefont {F.}~\bibnamefont {Schreck}},\ and\ \bibinfo {author} {\bibfnamefont {R.}~\bibnamefont {Grimm}},\ }\href {https://doi.org/10.1038/nature11065} {\bibfield  {journal} {\bibinfo  {journal} {Nature}\ }\textbf {\bibinfo {volume} {485}},\ \bibinfo {pages} {615} (\bibinfo {year} {2012})}\BibitemShut {NoStop}%
\bibitem [{\citenamefont {Cetina}\ \emph {et~al.}(2016)\citenamefont {Cetina}, \citenamefont {Jag}, \citenamefont {Lous}, \citenamefont {Fritsche}, \citenamefont {Walraven}, \citenamefont {Grimm}, \citenamefont {Levinsen}, \citenamefont {Parish}, \citenamefont {Schmidt}, \citenamefont {Knap},\ and\ \citenamefont {Demler}}]{cetina}%
  \BibitemOpen
  \bibfield  {author} {\bibinfo {author} {\bibfnamefont {M.}~\bibnamefont {Cetina}}, \bibinfo {author} {\bibfnamefont {M.}~\bibnamefont {Jag}}, \bibinfo {author} {\bibfnamefont {R.~S.}\ \bibnamefont {Lous}}, \bibinfo {author} {\bibfnamefont {I.}~\bibnamefont {Fritsche}}, \bibinfo {author} {\bibfnamefont {J.~T.~M.}\ \bibnamefont {Walraven}}, \bibinfo {author} {\bibfnamefont {R.}~\bibnamefont {Grimm}}, \bibinfo {author} {\bibfnamefont {J.}~\bibnamefont {Levinsen}}, \bibinfo {author} {\bibfnamefont {M.~M.}\ \bibnamefont {Parish}}, \bibinfo {author} {\bibfnamefont {R.}~\bibnamefont {Schmidt}}, \bibinfo {author} {\bibfnamefont {M.}~\bibnamefont {Knap}},\ and\ \bibinfo {author} {\bibfnamefont {E.}~\bibnamefont {Demler}},\ }\href {https://doi.org/10.1126/science.aaf5134} {\bibfield  {journal} {\bibinfo  {journal} {Science}\ }\textbf {\bibinfo {volume} {354}},\ \bibinfo {pages} {96} (\bibinfo {year} {2016})}\BibitemShut {NoStop}%
\bibitem [{\citenamefont {Jo}\ \emph {et~al.}(2009)\citenamefont {Jo}, \citenamefont {Lee}, \citenamefont {Choi}, \citenamefont {Christensen}, \citenamefont {Kim}, \citenamefont {Thywissen}, \citenamefont {Pritchard},\ and\ \citenamefont {Ketterle}}]{ferro2009}%
  \BibitemOpen
  \bibfield  {author} {\bibinfo {author} {\bibfnamefont {G.-B.}\ \bibnamefont {Jo}}, \bibinfo {author} {\bibfnamefont {Y.-R.}\ \bibnamefont {Lee}}, \bibinfo {author} {\bibfnamefont {J.-H.}\ \bibnamefont {Choi}}, \bibinfo {author} {\bibfnamefont {C.~A.}\ \bibnamefont {Christensen}}, \bibinfo {author} {\bibfnamefont {T.~H.}\ \bibnamefont {Kim}}, \bibinfo {author} {\bibfnamefont {J.~H.}\ \bibnamefont {Thywissen}}, \bibinfo {author} {\bibfnamefont {D.~E.}\ \bibnamefont {Pritchard}},\ and\ \bibinfo {author} {\bibfnamefont {W.}~\bibnamefont {Ketterle}},\ }\href {https://doi.org/10.1126/science.1177112} {\bibfield  {journal} {\bibinfo  {journal} {Science}\ }\textbf {\bibinfo {volume} {325}},\ \bibinfo {pages} {1521} (\bibinfo {year} {2009})}\BibitemShut {NoStop}%
\bibitem [{\citenamefont {Bresolin}\ \emph {et~al.}(2023)\citenamefont {Bresolin}, \citenamefont {Roy}, \citenamefont {Ferrari}, \citenamefont {Recati},\ and\ \citenamefont {Pavloff}}]{bresolin2023oscillating}%
  \BibitemOpen
  \bibfield  {author} {\bibinfo {author} {\bibfnamefont {S.}~\bibnamefont {Bresolin}}, \bibinfo {author} {\bibfnamefont {A.}~\bibnamefont {Roy}}, \bibinfo {author} {\bibfnamefont {G.}~\bibnamefont {Ferrari}}, \bibinfo {author} {\bibfnamefont {A.}~\bibnamefont {Recati}},\ and\ \bibinfo {author} {\bibfnamefont {N.}~\bibnamefont {Pavloff}},\ }\href {https://doi.org/10.1103/PhysRevLett.130.220403} {\bibfield  {journal} {\bibinfo  {journal} {Physical Review Letters}\ }\textbf {\bibinfo {volume} {130}},\ \bibinfo {pages} {220403} (\bibinfo {year} {2023})}\BibitemShut {NoStop}%
\bibitem [{\citenamefont {Tanzi}\ \emph {et~al.}(2019)\citenamefont {Tanzi}, \citenamefont {Lucioni}, \citenamefont {Fam\`a}, \citenamefont {Catani}, \citenamefont {Fioretti}, \citenamefont {Gabbanini}, \citenamefont {Bisset}, \citenamefont {Santos},\ and\ \citenamefont {Modugno}}]{PhysRevLett.122.130405}%
  \BibitemOpen
  \bibfield  {author} {\bibinfo {author} {\bibfnamefont {L.}~\bibnamefont {Tanzi}}, \bibinfo {author} {\bibfnamefont {E.}~\bibnamefont {Lucioni}}, \bibinfo {author} {\bibfnamefont {F.}~\bibnamefont {Fam\`a}}, \bibinfo {author} {\bibfnamefont {J.}~\bibnamefont {Catani}}, \bibinfo {author} {\bibfnamefont {A.}~\bibnamefont {Fioretti}}, \bibinfo {author} {\bibfnamefont {C.}~\bibnamefont {Gabbanini}}, \bibinfo {author} {\bibfnamefont {R.~N.}\ \bibnamefont {Bisset}}, \bibinfo {author} {\bibfnamefont {L.}~\bibnamefont {Santos}},\ and\ \bibinfo {author} {\bibfnamefont {G.}~\bibnamefont {Modugno}},\ }\href {https://doi.org/10.1103/PhysRevLett.122.130405} {\bibfield  {journal} {\bibinfo  {journal} {Physical Review Letters}\ }\textbf {\bibinfo {volume} {122}},\ \bibinfo {pages} {130405} (\bibinfo {year} {2019})}\BibitemShut {NoStop}%
\bibitem [{\citenamefont {B\"ottcher}\ \emph {et~al.}(2019)\citenamefont {B\"ottcher}, \citenamefont {Schmidt}, \citenamefont {Wenzel}, \citenamefont {Hertkorn}, \citenamefont {Guo}, \citenamefont {Langen},\ and\ \citenamefont {Pfau}}]{PhysRevX.9.011051}%
  \BibitemOpen
  \bibfield  {author} {\bibinfo {author} {\bibfnamefont {F.}~\bibnamefont {B\"ottcher}}, \bibinfo {author} {\bibfnamefont {J.-N.}\ \bibnamefont {Schmidt}}, \bibinfo {author} {\bibfnamefont {M.}~\bibnamefont {Wenzel}}, \bibinfo {author} {\bibfnamefont {J.}~\bibnamefont {Hertkorn}}, \bibinfo {author} {\bibfnamefont {M.}~\bibnamefont {Guo}}, \bibinfo {author} {\bibfnamefont {T.}~\bibnamefont {Langen}},\ and\ \bibinfo {author} {\bibfnamefont {T.}~\bibnamefont {Pfau}},\ }\href {https://doi.org/10.1103/PhysRevX.9.011051} {\bibfield  {journal} {\bibinfo  {journal} {Physical Review X}\ }\textbf {\bibinfo {volume} {9}},\ \bibinfo {pages} {011051} (\bibinfo {year} {2019})}\BibitemShut {NoStop}%
\bibitem [{\citenamefont {Chomaz}\ \emph {et~al.}(2019)\citenamefont {Chomaz}, \citenamefont {Petter}, \citenamefont {Ilzh\"ofer}, \citenamefont {Natale}, \citenamefont {Trautmann}, \citenamefont {Politi}, \citenamefont {Durastante}, \citenamefont {van Bijnen}, \citenamefont {Patscheider}, \citenamefont {Sohmen}, \citenamefont {Mark},\ and\ \citenamefont {Ferlaino}}]{PhysRevX.9.021012}%
  \BibitemOpen
  \bibfield  {author} {\bibinfo {author} {\bibfnamefont {L.}~\bibnamefont {Chomaz}}, \bibinfo {author} {\bibfnamefont {D.}~\bibnamefont {Petter}}, \bibinfo {author} {\bibfnamefont {P.}~\bibnamefont {Ilzh\"ofer}}, \bibinfo {author} {\bibfnamefont {G.}~\bibnamefont {Natale}}, \bibinfo {author} {\bibfnamefont {A.}~\bibnamefont {Trautmann}}, \bibinfo {author} {\bibfnamefont {C.}~\bibnamefont {Politi}}, \bibinfo {author} {\bibfnamefont {G.}~\bibnamefont {Durastante}}, \bibinfo {author} {\bibfnamefont {R.~M.~W.}\ \bibnamefont {van Bijnen}}, \bibinfo {author} {\bibfnamefont {A.}~\bibnamefont {Patscheider}}, \bibinfo {author} {\bibfnamefont {M.}~\bibnamefont {Sohmen}}, \bibinfo {author} {\bibfnamefont {M.~J.}\ \bibnamefont {Mark}},\ and\ \bibinfo {author} {\bibfnamefont {F.}~\bibnamefont {Ferlaino}},\ }\href {https://doi.org/10.1103/PhysRevX.9.021012} {\bibfield  {journal} {\bibinfo  {journal} {Physical Review X}\ }\textbf {\bibinfo {volume} {9}},\ \bibinfo {pages} {021012} (\bibinfo {year} {2019})}\BibitemShut
  {NoStop}%
\bibitem [{\citenamefont {Schellekens}\ \emph {et~al.}(2005)\citenamefont {Schellekens}, \citenamefont {Hoppeler}, \citenamefont {Perrin}, \citenamefont {Gomes}, \citenamefont {Boiron}, \citenamefont {Aspect},\ and\ \citenamefont {Westbrook}}]{schellekens2005hanbury}%
  \BibitemOpen
  \bibfield  {author} {\bibinfo {author} {\bibfnamefont {M.}~\bibnamefont {Schellekens}}, \bibinfo {author} {\bibfnamefont {R.}~\bibnamefont {Hoppeler}}, \bibinfo {author} {\bibfnamefont {A.}~\bibnamefont {Perrin}}, \bibinfo {author} {\bibfnamefont {J.~V.}\ \bibnamefont {Gomes}}, \bibinfo {author} {\bibfnamefont {D.}~\bibnamefont {Boiron}}, \bibinfo {author} {\bibfnamefont {A.}~\bibnamefont {Aspect}},\ and\ \bibinfo {author} {\bibfnamefont {C.~I.}\ \bibnamefont {Westbrook}},\ }\href {https://doi.org/10.1126/science.1118024} {\bibfield  {journal} {\bibinfo  {journal} {Science}\ }\textbf {\bibinfo {volume} {310}},\ \bibinfo {pages} {648} (\bibinfo {year} {2005})}\BibitemShut {NoStop}%
\bibitem [{\citenamefont {Jeltes}\ \emph {et~al.}(2007)\citenamefont {Jeltes}, \citenamefont {McNamara}, \citenamefont {Hogervorst}, \citenamefont {Vassen}, \citenamefont {Krachmalnicoff}, \citenamefont {Schellekens}, \citenamefont {Perrin}, \citenamefont {Chang}, \citenamefont {Boiron}, \citenamefont {Aspect},\ and\ \citenamefont {Westbrook}}]{jeltes2007comparison}%
  \BibitemOpen
  \bibfield  {author} {\bibinfo {author} {\bibfnamefont {T.}~\bibnamefont {Jeltes}}, \bibinfo {author} {\bibfnamefont {J.~M.}\ \bibnamefont {McNamara}}, \bibinfo {author} {\bibfnamefont {W.}~\bibnamefont {Hogervorst}}, \bibinfo {author} {\bibfnamefont {W.}~\bibnamefont {Vassen}}, \bibinfo {author} {\bibfnamefont {V.}~\bibnamefont {Krachmalnicoff}}, \bibinfo {author} {\bibfnamefont {M.}~\bibnamefont {Schellekens}}, \bibinfo {author} {\bibfnamefont {A.}~\bibnamefont {Perrin}}, \bibinfo {author} {\bibfnamefont {H.}~\bibnamefont {Chang}}, \bibinfo {author} {\bibfnamefont {D.}~\bibnamefont {Boiron}}, \bibinfo {author} {\bibfnamefont {A.}~\bibnamefont {Aspect}},\ and\ \bibinfo {author} {\bibfnamefont {C.~I.}\ \bibnamefont {Westbrook}},\ }\href {https://doi.org/https://doi.org/10.1038/nature05513} {\bibfield  {journal} {\bibinfo  {journal} {Nature}\ }\textbf {\bibinfo {volume} {445}},\ \bibinfo {pages} {402} (\bibinfo {year} {2007})}\BibitemShut {NoStop}%
\bibitem [{\citenamefont {Hodgman}\ \emph {et~al.}(2011)\citenamefont {Hodgman}, \citenamefont {Dall}, \citenamefont {Manning}, \citenamefont {Baldwin},\ and\ \citenamefont {Truscott}}]{hodgman2011direct}%
  \BibitemOpen
  \bibfield  {author} {\bibinfo {author} {\bibfnamefont {S.~S.}\ \bibnamefont {Hodgman}}, \bibinfo {author} {\bibfnamefont {R.~G.}\ \bibnamefont {Dall}}, \bibinfo {author} {\bibfnamefont {A.~G.}\ \bibnamefont {Manning}}, \bibinfo {author} {\bibfnamefont {K.~G.~H.}\ \bibnamefont {Baldwin}},\ and\ \bibinfo {author} {\bibfnamefont {A.~G.}\ \bibnamefont {Truscott}},\ }\href {https://doi.org/https://doi.org/10.1126/science.1198481} {\bibfield  {journal} {\bibinfo  {journal} {Science}\ }\textbf {\bibinfo {volume} {331}},\ \bibinfo {pages} {1046} (\bibinfo {year} {2011})}\BibitemShut {NoStop}%
\bibitem [{\citenamefont {Manning}\ \emph {et~al.}(2015)\citenamefont {Manning}, \citenamefont {Khakimov}, \citenamefont {Dall},\ and\ \citenamefont {Truscott}}]{manning2015wheeler}%
  \BibitemOpen
  \bibfield  {author} {\bibinfo {author} {\bibfnamefont {A.~G.}\ \bibnamefont {Manning}}, \bibinfo {author} {\bibfnamefont {R.~I.}\ \bibnamefont {Khakimov}}, \bibinfo {author} {\bibfnamefont {R.~G.}\ \bibnamefont {Dall}},\ and\ \bibinfo {author} {\bibfnamefont {A.~G.}\ \bibnamefont {Truscott}},\ }\href {https://doi.org/https://doi.org/10.1038/nphys3343} {\bibfield  {journal} {\bibinfo  {journal} {Nature Physics}\ }\textbf {\bibinfo {volume} {11}},\ \bibinfo {pages} {539} (\bibinfo {year} {2015})}\BibitemShut {NoStop}%
\bibitem [{\citenamefont {Lopes}\ \emph {et~al.}(2015)\citenamefont {Lopes}, \citenamefont {Imanaliev}, \citenamefont {Aspect}, \citenamefont {Cheneau}, \citenamefont {Boiron},\ and\ \citenamefont {Westbrook}}]{lopes2015atomic}%
  \BibitemOpen
  \bibfield  {author} {\bibinfo {author} {\bibfnamefont {R.}~\bibnamefont {Lopes}}, \bibinfo {author} {\bibfnamefont {A.}~\bibnamefont {Imanaliev}}, \bibinfo {author} {\bibfnamefont {A.}~\bibnamefont {Aspect}}, \bibinfo {author} {\bibfnamefont {M.}~\bibnamefont {Cheneau}}, \bibinfo {author} {\bibfnamefont {D.}~\bibnamefont {Boiron}},\ and\ \bibinfo {author} {\bibfnamefont {C.~I.}\ \bibnamefont {Westbrook}},\ }\href {https://doi.org/https://doi.org/10.1038/nphys3343} {\bibfield  {journal} {\bibinfo  {journal} {Nature}\ }\textbf {\bibinfo {volume} {520}},\ \bibinfo {pages} {66} (\bibinfo {year} {2015})}\BibitemShut {NoStop}%
\bibitem [{\citenamefont {Perrin}\ \emph {et~al.}(2007)\citenamefont {Perrin}, \citenamefont {Chang}, \citenamefont {Krachmalnicoff}, \citenamefont {Schellekens}, \citenamefont {Boiron}, \citenamefont {Aspect},\ and\ \citenamefont {Westbrook}}]{PhysRevLett.99.150405}%
  \BibitemOpen
  \bibfield  {author} {\bibinfo {author} {\bibfnamefont {A.}~\bibnamefont {Perrin}}, \bibinfo {author} {\bibfnamefont {H.}~\bibnamefont {Chang}}, \bibinfo {author} {\bibfnamefont {V.}~\bibnamefont {Krachmalnicoff}}, \bibinfo {author} {\bibfnamefont {M.}~\bibnamefont {Schellekens}}, \bibinfo {author} {\bibfnamefont {D.}~\bibnamefont {Boiron}}, \bibinfo {author} {\bibfnamefont {A.}~\bibnamefont {Aspect}},\ and\ \bibinfo {author} {\bibfnamefont {C.~I.}\ \bibnamefont {Westbrook}},\ }\href {https://doi.org/10.1103/PhysRevLett.99.150405} {\bibfield  {journal} {\bibinfo  {journal} {Physical Review Letters}\ }\textbf {\bibinfo {volume} {99}},\ \bibinfo {pages} {150405} (\bibinfo {year} {2007})}\BibitemShut {NoStop}%
\bibitem [{\citenamefont {Khakimov}\ \emph {et~al.}(2016)\citenamefont {Khakimov}, \citenamefont {Henson}, \citenamefont {Shin}, \citenamefont {Hodgman}, \citenamefont {Dall}, \citenamefont {Baldwin},\ and\ \citenamefont {Truscott}}]{khakimov2016ghost}%
  \BibitemOpen
  \bibfield  {author} {\bibinfo {author} {\bibfnamefont {R.~I.}\ \bibnamefont {Khakimov}}, \bibinfo {author} {\bibfnamefont {B.~M.}\ \bibnamefont {Henson}}, \bibinfo {author} {\bibfnamefont {D.~K.}\ \bibnamefont {Shin}}, \bibinfo {author} {\bibfnamefont {S.~S.}\ \bibnamefont {Hodgman}}, \bibinfo {author} {\bibfnamefont {R.~G.}\ \bibnamefont {Dall}}, \bibinfo {author} {\bibfnamefont {K.~G.~H.}\ \bibnamefont {Baldwin}},\ and\ \bibinfo {author} {\bibfnamefont {A.~G.}\ \bibnamefont {Truscott}},\ }\href {https://doi.org/https://doi.org/10.1038/nphys3343} {\bibfield  {journal} {\bibinfo  {journal} {Nature}\ }\textbf {\bibinfo {volume} {540}},\ \bibinfo {pages} {100} (\bibinfo {year} {2016})}\BibitemShut {NoStop}%
\bibitem [{\citenamefont {Hodgman}\ \emph {et~al.}(2017)\citenamefont {Hodgman}, \citenamefont {Khakimov}, \citenamefont {Lewis-Swan}, \citenamefont {Truscott},\ and\ \citenamefont {Kheruntsyan}}]{hodgman2017solving}%
  \BibitemOpen
  \bibfield  {author} {\bibinfo {author} {\bibfnamefont {S.~S.}\ \bibnamefont {Hodgman}}, \bibinfo {author} {\bibfnamefont {R.~I.}\ \bibnamefont {Khakimov}}, \bibinfo {author} {\bibfnamefont {R.~J.}\ \bibnamefont {Lewis-Swan}}, \bibinfo {author} {\bibfnamefont {A.~G.}\ \bibnamefont {Truscott}},\ and\ \bibinfo {author} {\bibfnamefont {K.~V.}\ \bibnamefont {Kheruntsyan}},\ }\href {https://doi.org/10.1103/PhysRevLett.118.240402} {\bibfield  {journal} {\bibinfo  {journal} {Physical Review Letters}\ }\textbf {\bibinfo {volume} {118}},\ \bibinfo {pages} {240402} (\bibinfo {year} {2017})}\BibitemShut {NoStop}%
\bibitem [{\citenamefont {Thomas}\ \emph {et~al.}(2022)\citenamefont {Thomas}, \citenamefont {Henson}, \citenamefont {Wang}, \citenamefont {Lewis-Swan}, \citenamefont {Kheruntsyan}, \citenamefont {Hodgman},\ and\ \citenamefont {Truscott}}]{thomas2022matter}%
  \BibitemOpen
  \bibfield  {author} {\bibinfo {author} {\bibfnamefont {K.~F.}\ \bibnamefont {Thomas}}, \bibinfo {author} {\bibfnamefont {B.~M.}\ \bibnamefont {Henson}}, \bibinfo {author} {\bibfnamefont {Y.}~\bibnamefont {Wang}}, \bibinfo {author} {\bibfnamefont {R.~J.}\ \bibnamefont {Lewis-Swan}}, \bibinfo {author} {\bibfnamefont {K.~V.}\ \bibnamefont {Kheruntsyan}}, \bibinfo {author} {\bibfnamefont {S.~S.}\ \bibnamefont {Hodgman}},\ and\ \bibinfo {author} {\bibfnamefont {A.~G.}\ \bibnamefont {Truscott}},\ }\href {https://doi.org/https://doi.org/10.1140/epjd/s10053-022-00551-y} {\bibfield  {journal} {\bibinfo  {journal} {The European Physical Journal D}\ }\textbf {\bibinfo {volume} {76}},\ \bibinfo {pages} {244} (\bibinfo {year} {2022})}\BibitemShut {NoStop}%
\bibitem [{\citenamefont {Fabbri}\ \emph {et~al.}(2015)\citenamefont {Fabbri}, \citenamefont {Panfil}, \citenamefont {Cl{\'e}ment}, \citenamefont {Fallani}, \citenamefont {Inguscio}, \citenamefont {Fort},\ and\ \citenamefont {Caux}}]{fabbri2015dynamical}%
  \BibitemOpen
  \bibfield  {author} {\bibinfo {author} {\bibfnamefont {N.}~\bibnamefont {Fabbri}}, \bibinfo {author} {\bibfnamefont {M.}~\bibnamefont {Panfil}}, \bibinfo {author} {\bibfnamefont {D.}~\bibnamefont {Cl{\'e}ment}}, \bibinfo {author} {\bibfnamefont {L.}~\bibnamefont {Fallani}}, \bibinfo {author} {\bibfnamefont {M.}~\bibnamefont {Inguscio}}, \bibinfo {author} {\bibfnamefont {C.}~\bibnamefont {Fort}},\ and\ \bibinfo {author} {\bibfnamefont {J.-S.}\ \bibnamefont {Caux}},\ }\href {https://doi.org/10.1103/PhysRevA.91.043617} {\bibfield  {journal} {\bibinfo  {journal} {Physical Review A}\ }\textbf {\bibinfo {volume} {91}},\ \bibinfo {pages} {043617} (\bibinfo {year} {2015})}\BibitemShut {NoStop}%
\bibitem [{\citenamefont {Tenart}\ \emph {et~al.}(2020)\citenamefont {Tenart}, \citenamefont {Carcy}, \citenamefont {Cayla}, \citenamefont {Bourdel}, \citenamefont {Mancini},\ and\ \citenamefont {Cl{\'e}ment}}]{tenart2020two}%
  \BibitemOpen
  \bibfield  {author} {\bibinfo {author} {\bibfnamefont {A.}~\bibnamefont {Tenart}}, \bibinfo {author} {\bibfnamefont {C.}~\bibnamefont {Carcy}}, \bibinfo {author} {\bibfnamefont {H.}~\bibnamefont {Cayla}}, \bibinfo {author} {\bibfnamefont {T.}~\bibnamefont {Bourdel}}, \bibinfo {author} {\bibfnamefont {M.}~\bibnamefont {Mancini}},\ and\ \bibinfo {author} {\bibfnamefont {D.}~\bibnamefont {Cl{\'e}ment}},\ }\href {https://doi.org/10.1103/PhysRevResearch.2.013017} {\bibfield  {journal} {\bibinfo  {journal} {Physical Review Research}\ }\textbf {\bibinfo {volume} {2}},\ \bibinfo {pages} {013017} (\bibinfo {year} {2020})}\BibitemShut {NoStop}%
\bibitem [{\citenamefont {Butera}\ \emph {et~al.}(2021)\citenamefont {Butera}, \citenamefont {Cl{\'e}ment},\ and\ \citenamefont {Carusotto}}]{butera2021position}%
  \BibitemOpen
  \bibfield  {author} {\bibinfo {author} {\bibfnamefont {S.}~\bibnamefont {Butera}}, \bibinfo {author} {\bibfnamefont {D.}~\bibnamefont {Cl{\'e}ment}},\ and\ \bibinfo {author} {\bibfnamefont {I.}~\bibnamefont {Carusotto}},\ }\href {https://doi.org/10.1103/PhysRevA.103.013302} {\bibfield  {journal} {\bibinfo  {journal} {Physical Review A}\ }\textbf {\bibinfo {volume} {103}},\ \bibinfo {pages} {013302} (\bibinfo {year} {2021})}\BibitemShut {NoStop}%
\bibitem [{\citenamefont {Herc{\'e}}\ \emph {et~al.}(2021)\citenamefont {Herc{\'e}}, \citenamefont {Carcy}, \citenamefont {Tenart}, \citenamefont {Bureik}, \citenamefont {Dareau}, \citenamefont {Cl{\'e}ment},\ and\ \citenamefont {Roscilde}}]{herce2021studying}%
  \BibitemOpen
  \bibfield  {author} {\bibinfo {author} {\bibfnamefont {G.}~\bibnamefont {Herc{\'e}}}, \bibinfo {author} {\bibfnamefont {C.}~\bibnamefont {Carcy}}, \bibinfo {author} {\bibfnamefont {A.}~\bibnamefont {Tenart}}, \bibinfo {author} {\bibfnamefont {J.-P.}\ \bibnamefont {Bureik}}, \bibinfo {author} {\bibfnamefont {A.}~\bibnamefont {Dareau}}, \bibinfo {author} {\bibfnamefont {D.}~\bibnamefont {Cl{\'e}ment}},\ and\ \bibinfo {author} {\bibfnamefont {T.}~\bibnamefont {Roscilde}},\ }\href {https://doi.org/10.1103/PhysRevA.104.L011301} {\bibfield  {journal} {\bibinfo  {journal} {Physical Review A}\ }\textbf {\bibinfo {volume} {104}},\ \bibinfo {pages} {L011301} (\bibinfo {year} {2021})}\BibitemShut {NoStop}%
\bibitem [{\citenamefont {McNamara}\ \emph {et~al.}(2006)\citenamefont {McNamara}, \citenamefont {Jeltes}, \citenamefont {Tychkov}, \citenamefont {Hogervorst},\ and\ \citenamefont {Vassen}}]{mcnamara2006degenerate}%
  \BibitemOpen
  \bibfield  {author} {\bibinfo {author} {\bibfnamefont {J.~M.}\ \bibnamefont {McNamara}}, \bibinfo {author} {\bibfnamefont {T.}~\bibnamefont {Jeltes}}, \bibinfo {author} {\bibfnamefont {A.~S.}\ \bibnamefont {Tychkov}}, \bibinfo {author} {\bibfnamefont {W.}~\bibnamefont {Hogervorst}},\ and\ \bibinfo {author} {\bibfnamefont {W.}~\bibnamefont {Vassen}},\ }\href {https://doi.org/10.1103/PhysRevLett.97.080404} {\bibfield  {journal} {\bibinfo  {journal} {Physical Review Letters}\ }\textbf {\bibinfo {volume} {97}},\ \bibinfo {pages} {080404} (\bibinfo {year} {2006})}\BibitemShut {NoStop}%
\bibitem [{\citenamefont {Hirsch}\ \emph {et~al.}(2021)\citenamefont {Hirsch}, \citenamefont {Cocks},\ and\ \citenamefont {Hodgman}}]{hirsch2021close}%
  \BibitemOpen
  \bibfield  {author} {\bibinfo {author} {\bibfnamefont {T.~M.~F.}\ \bibnamefont {Hirsch}}, \bibinfo {author} {\bibfnamefont {D.~G.}\ \bibnamefont {Cocks}},\ and\ \bibinfo {author} {\bibfnamefont {S.~S.}\ \bibnamefont {Hodgman}},\ }\href {https://doi.org/10.1103/PhysRevA.104.033317} {\bibfield  {journal} {\bibinfo  {journal} {Physical Review A}\ }\textbf {\bibinfo {volume} {104}},\ \bibinfo {pages} {033317} (\bibinfo {year} {2021})}\BibitemShut {NoStop}%
\bibitem [{\citenamefont {Goosen}\ \emph {et~al.}(2010)\citenamefont {Goosen}, \citenamefont {Tiecke}, \citenamefont {Vassen},\ and\ \citenamefont {Kokkelmans}}]{PhysRevA.82.042713}%
  \BibitemOpen
  \bibfield  {author} {\bibinfo {author} {\bibfnamefont {M.~R.}\ \bibnamefont {Goosen}}, \bibinfo {author} {\bibfnamefont {T.~G.}\ \bibnamefont {Tiecke}}, \bibinfo {author} {\bibfnamefont {W.}~\bibnamefont {Vassen}},\ and\ \bibinfo {author} {\bibfnamefont {S.~J. J. M.~F.}\ \bibnamefont {Kokkelmans}},\ }\href {https://doi.org/10.1103/PhysRevA.82.042713} {\bibfield  {journal} {\bibinfo  {journal} {Physical Review A}\ }\textbf {\bibinfo {volume} {82}},\ \bibinfo {pages} {042713} (\bibinfo {year} {2010})}\BibitemShut {NoStop}%
\bibitem [{\citenamefont {Thomas}\ \emph {et~al.}(2023)\citenamefont {Thomas}, \citenamefont {Ou}, \citenamefont {Henson}, \citenamefont {Baiju}, \citenamefont {Hodgman},\ and\ \citenamefont {Truscott}}]{thomas2023production}%
  \BibitemOpen
  \bibfield  {author} {\bibinfo {author} {\bibfnamefont {K.~F.}\ \bibnamefont {Thomas}}, \bibinfo {author} {\bibfnamefont {Z.}~\bibnamefont {Ou}}, \bibinfo {author} {\bibfnamefont {B.~M.}\ \bibnamefont {Henson}}, \bibinfo {author} {\bibfnamefont {A.~A.}\ \bibnamefont {Baiju}}, \bibinfo {author} {\bibfnamefont {S.~S.}\ \bibnamefont {Hodgman}},\ and\ \bibinfo {author} {\bibfnamefont {A.~G.}\ \bibnamefont {Truscott}},\ }\href {https://doi.org/10.1103/PhysRevA.107.033313} {\bibfield  {journal} {\bibinfo  {journal} {Physical Review A}\ }\textbf {\bibinfo {volume} {107}},\ \bibinfo {pages} {033313} (\bibinfo {year} {2023})}\BibitemShut {NoStop}%
\bibitem [{\citenamefont {Manning}\ \emph {et~al.}(2010)\citenamefont {Manning}, \citenamefont {Hodgman}, \citenamefont {Dall}, \citenamefont {Johnsson},\ and\ \citenamefont {Truscott}}]{manning2010hanbury}%
  \BibitemOpen
  \bibfield  {author} {\bibinfo {author} {\bibfnamefont {A.~G.}\ \bibnamefont {Manning}}, \bibinfo {author} {\bibfnamefont {S.~S.}\ \bibnamefont {Hodgman}}, \bibinfo {author} {\bibfnamefont {R.~G.}\ \bibnamefont {Dall}}, \bibinfo {author} {\bibfnamefont {M.~T.}\ \bibnamefont {Johnsson}},\ and\ \bibinfo {author} {\bibfnamefont {A.~G.}\ \bibnamefont {Truscott}},\ }\href {https://doi.org/https://doi.org/10.1364/OE.18.018712} {\bibfield  {journal} {\bibinfo  {journal} {Optics Express}\ }\textbf {\bibinfo {volume} {18}},\ \bibinfo {pages} {18712} (\bibinfo {year} {2010})}\BibitemShut {NoStop}%
\bibitem [{\citenamefont {Henson}\ \emph {et~al.}(2022{\natexlab{a}})\citenamefont {Henson}, \citenamefont {Ross}, \citenamefont {Thomas}, \citenamefont {Kuhn}, \citenamefont {Shin}, \citenamefont {Hodgman}, \citenamefont {Zhang}, \citenamefont {Tang}, \citenamefont {Drake}, \citenamefont {Bondy}, \citenamefont {Truscott},\ and\ \citenamefont {Baldwin}}]{henson2022measurement}%
  \BibitemOpen
  \bibfield  {author} {\bibinfo {author} {\bibfnamefont {B.~M.}\ \bibnamefont {Henson}}, \bibinfo {author} {\bibfnamefont {J.~A.}\ \bibnamefont {Ross}}, \bibinfo {author} {\bibfnamefont {K.~F.}\ \bibnamefont {Thomas}}, \bibinfo {author} {\bibfnamefont {C.~N.}\ \bibnamefont {Kuhn}}, \bibinfo {author} {\bibfnamefont {D.~K.}\ \bibnamefont {Shin}}, \bibinfo {author} {\bibfnamefont {S.~S.}\ \bibnamefont {Hodgman}}, \bibinfo {author} {\bibfnamefont {Y.-H.}\ \bibnamefont {Zhang}}, \bibinfo {author} {\bibfnamefont {L.-Y.}\ \bibnamefont {Tang}}, \bibinfo {author} {\bibfnamefont {G.~W.~F.}\ \bibnamefont {Drake}}, \bibinfo {author} {\bibfnamefont {A.~T.}\ \bibnamefont {Bondy}}, \bibinfo {author} {\bibfnamefont {A.~G.}\ \bibnamefont {Truscott}},\ and\ \bibinfo {author} {\bibfnamefont {K.~G.~H.}\ \bibnamefont {Baldwin}},\ }\href {https://doi.org/10.1126/science.abk2502} {\bibfield  {journal} {\bibinfo  {journal} {Science}\ }\textbf {\bibinfo {volume} {376}},\ \bibinfo {pages} {199} (\bibinfo {year}
  {2022}{\natexlab{a}})}\BibitemShut {NoStop}%
\bibitem [{\citenamefont {Henson}\ \emph {et~al.}(2022{\natexlab{b}})\citenamefont {Henson}, \citenamefont {Thomas}, \citenamefont {Mehdi}, \citenamefont {Burnett}, \citenamefont {Ross}, \citenamefont {Hodgman},\ and\ \citenamefont {Truscott}}]{henson2022trap}%
  \BibitemOpen
  \bibfield  {author} {\bibinfo {author} {\bibfnamefont {B.~M.}\ \bibnamefont {Henson}}, \bibinfo {author} {\bibfnamefont {K.~F.}\ \bibnamefont {Thomas}}, \bibinfo {author} {\bibfnamefont {Z.}~\bibnamefont {Mehdi}}, \bibinfo {author} {\bibfnamefont {T.~G.}\ \bibnamefont {Burnett}}, \bibinfo {author} {\bibfnamefont {J.~A.}\ \bibnamefont {Ross}}, \bibinfo {author} {\bibfnamefont {S.~S.}\ \bibnamefont {Hodgman}},\ and\ \bibinfo {author} {\bibfnamefont {A.~G.}\ \bibnamefont {Truscott}},\ }\href {https://doi.org/https://doi.org/10.1364/OE.455009} {\bibfield  {journal} {\bibinfo  {journal} {Optics Express}\ }\textbf {\bibinfo {volume} {30}},\ \bibinfo {pages} {13252} (\bibinfo {year} {2022}{\natexlab{b}})}\BibitemShut {NoStop}%
\bibitem [{\citenamefont {Dall}\ and\ \citenamefont {Truscott}(2007)}]{dall2007bose}%
  \BibitemOpen
  \bibfield  {author} {\bibinfo {author} {\bibfnamefont {R.~G.}\ \bibnamefont {Dall}}\ and\ \bibinfo {author} {\bibfnamefont {A.~G.}\ \bibnamefont {Truscott}},\ }\href {https://doi.org/https://doi.org/10.1016/j.optcom.2006.09.031} {\bibfield  {journal} {\bibinfo  {journal} {Optics Communications}\ }\textbf {\bibinfo {volume} {270}},\ \bibinfo {pages} {255} (\bibinfo {year} {2007})}\BibitemShut {NoStop}%
\bibitem [{\citenamefont {Henson}\ \emph {et~al.}(2018)\citenamefont {Henson}, \citenamefont {Yue}, \citenamefont {Hodgman}, \citenamefont {Shin}, \citenamefont {Smirnov}, \citenamefont {Ostrovskaya}, \citenamefont {Guan},\ and\ \citenamefont {Truscott}}]{PhysRevA.97.063601}%
  \BibitemOpen
  \bibfield  {author} {\bibinfo {author} {\bibfnamefont {B.~M.}\ \bibnamefont {Henson}}, \bibinfo {author} {\bibfnamefont {X.}~\bibnamefont {Yue}}, \bibinfo {author} {\bibfnamefont {S.~S.}\ \bibnamefont {Hodgman}}, \bibinfo {author} {\bibfnamefont {D.~K.}\ \bibnamefont {Shin}}, \bibinfo {author} {\bibfnamefont {L.~A.}\ \bibnamefont {Smirnov}}, \bibinfo {author} {\bibfnamefont {E.~A.}\ \bibnamefont {Ostrovskaya}}, \bibinfo {author} {\bibfnamefont {X.~W.}\ \bibnamefont {Guan}},\ and\ \bibinfo {author} {\bibfnamefont {A.~G.}\ \bibnamefont {Truscott}},\ }\href {https://doi.org/10.1103/PhysRevA.97.063601} {\bibfield  {journal} {\bibinfo  {journal} {Physical Review A}\ }\textbf {\bibinfo {volume} {97}},\ \bibinfo {pages} {063601} (\bibinfo {year} {2018})}\BibitemShut {NoStop}%
\bibitem [{\citenamefont {Fedichev}\ \emph {et~al.}(1998)\citenamefont {Fedichev}, \citenamefont {Shlyapnikov},\ and\ \citenamefont {Walraven}}]{fedichev1998damping}%
  \BibitemOpen
  \bibfield  {author} {\bibinfo {author} {\bibfnamefont {P.~O.}\ \bibnamefont {Fedichev}}, \bibinfo {author} {\bibfnamefont {G.~V.}\ \bibnamefont {Shlyapnikov}},\ and\ \bibinfo {author} {\bibfnamefont {J.~T.~M.}\ \bibnamefont {Walraven}},\ }\href {https://doi.org/10.1103/PhysRevLett.80.2269} {\bibfield  {journal} {\bibinfo  {journal} {Physical Review Letters}\ }\textbf {\bibinfo {volume} {80}},\ \bibinfo {pages} {2269} (\bibinfo {year} {1998})}\BibitemShut {NoStop}%
\bibitem [{\citenamefont {Ferrari}\ \emph {et~al.}(2002)\citenamefont {Ferrari}, \citenamefont {Inguscio}, \citenamefont {Jastrzebski}, \citenamefont {Modugno}, \citenamefont {Roati},\ and\ \citenamefont {Simoni}}]{ferrari2002collisional}%
  \BibitemOpen
  \bibfield  {author} {\bibinfo {author} {\bibfnamefont {G.}~\bibnamefont {Ferrari}}, \bibinfo {author} {\bibfnamefont {M.}~\bibnamefont {Inguscio}}, \bibinfo {author} {\bibfnamefont {W.}~\bibnamefont {Jastrzebski}}, \bibinfo {author} {\bibfnamefont {G.}~\bibnamefont {Modugno}}, \bibinfo {author} {\bibfnamefont {G.}~\bibnamefont {Roati}},\ and\ \bibinfo {author} {\bibfnamefont {A.}~\bibnamefont {Simoni}},\ }\href {https://doi.org/10.1103/PhysRevLett.89.053202} {\bibfield  {journal} {\bibinfo  {journal} {Physical Review Letters}\ }\textbf {\bibinfo {volume} {89}},\ \bibinfo {pages} {053202} (\bibinfo {year} {2002})}\BibitemShut {NoStop}%
\bibitem [{\citenamefont {Jaskula}\ \emph {et~al.}(2010)\citenamefont {Jaskula}, \citenamefont {Bonneau}, \citenamefont {Partridge}, \citenamefont {Krachmalnicoff}, \citenamefont {Deuar}, \citenamefont {Kheruntsyan}, \citenamefont {Aspect}, \citenamefont {Boiron},\ and\ \citenamefont {Westbrook}}]{PhysRevLett.105.190402}%
  \BibitemOpen
  \bibfield  {author} {\bibinfo {author} {\bibfnamefont {J.-C.}\ \bibnamefont {Jaskula}}, \bibinfo {author} {\bibfnamefont {M.}~\bibnamefont {Bonneau}}, \bibinfo {author} {\bibfnamefont {G.~B.}\ \bibnamefont {Partridge}}, \bibinfo {author} {\bibfnamefont {V.}~\bibnamefont {Krachmalnicoff}}, \bibinfo {author} {\bibfnamefont {P.}~\bibnamefont {Deuar}}, \bibinfo {author} {\bibfnamefont {K.~V.}\ \bibnamefont {Kheruntsyan}}, \bibinfo {author} {\bibfnamefont {A.}~\bibnamefont {Aspect}}, \bibinfo {author} {\bibfnamefont {D.}~\bibnamefont {Boiron}},\ and\ \bibinfo {author} {\bibfnamefont {C.~I.}\ \bibnamefont {Westbrook}},\ }\href {https://doi.org/10.1103/PhysRevLett.105.190402} {\bibfield  {journal} {\bibinfo  {journal} {Physical Review Letters}\ }\textbf {\bibinfo {volume} {105}},\ \bibinfo {pages} {190402} (\bibinfo {year} {2010})}\BibitemShut {NoStop}%
\bibitem [{\citenamefont {Lewis-Swan}\ and\ \citenamefont {Kheruntsyan}(2015)}]{PhysRevA.91.052114}%
  \BibitemOpen
  \bibfield  {author} {\bibinfo {author} {\bibfnamefont {R.~J.}\ \bibnamefont {Lewis-Swan}}\ and\ \bibinfo {author} {\bibfnamefont {K.~V.}\ \bibnamefont {Kheruntsyan}},\ }\href {https://doi.org/10.1103/PhysRevA.91.052114} {\bibfield  {journal} {\bibinfo  {journal} {Physical Review A}\ }\textbf {\bibinfo {volume} {91}},\ \bibinfo {pages} {052114} (\bibinfo {year} {2015})}\BibitemShut {NoStop}%
\bibitem [{\citenamefont {Geiger}\ and\ \citenamefont {Trupke}(2018)}]{PhysRevLett.120.043602}%
  \BibitemOpen
  \bibfield  {author} {\bibinfo {author} {\bibfnamefont {R.}~\bibnamefont {Geiger}}\ and\ \bibinfo {author} {\bibfnamefont {M.}~\bibnamefont {Trupke}},\ }\href {https://doi.org/10.1103/PhysRevLett.120.043602} {\bibfield  {journal} {\bibinfo  {journal} {Physical Review Letters}\ }\textbf {\bibinfo {volume} {120}},\ \bibinfo {pages} {043602} (\bibinfo {year} {2018})}\BibitemShut {NoStop}%
\end{thebibliography}%

\end{document}